\documentclass[11pt,a4paper]{article}
\pdfoutput=1
\usepackage{jheppub}
\usepackage{tikz}
\usepackage{array}
\usepackage{MnSymbol}
\usepackage{dsfont}
\usepackage{slashed}
\usepackage{mathtools}
\allowdisplaybreaks
\usepackage{mathrsfs}
\usepackage[squaren,Gray]{SIunits}
\usepackage{booktabs}
\usepackage{appendix}
\newcommand{\overbar}[1]{\mkern 1.5mu\overline{\mkern-1.5mu#1\mkern-1.5mu}\mkern 1.5mu}

\newcommand{\MSbar}{\overbar{\text{MS}}}
\newcommand{\RI}{\text{RI}^\prime\mkern-3mu\raisebox{1pt}{$/$}\text{SMOM}}
\newcommand{\SU}[1]{\ensuremath{\text{SU}(#1)}}
\newcommand{\cO}{\mathcal{O}}%

\newcommand{\mcemd}{\multicolumn{1}{c}{---}}

\newlength\B
\newcommand*{\tikzperp}{\begin{tikzpicture}%
\draw[line width=0.1\B,cap=round](0\B,0.05\B)--(0\B,0.95\B);
\draw[line width=0.1\B,cap=round](-0.375\B,0.05\B)--(+0.375\B,0.05\B);
\end{tikzpicture}}%
\newcommand*{\tikzparallel}{\begin{tikzpicture}%
\draw[line width=0.1\B,cap=round](-0.125\B,0.05\B)--(-0.125\B,0.95\B);
\draw[line width=0.1\B,cap=round](+0.125\B,0.05\B)--(+0.125\B,0.95\B);
\end{tikzpicture}}%
\newcommand{\goodsymbol}[2]{\settoheight{\B}{$#1B$}#2}
\newcommand{\goodperp}{{\mathpalette\goodsymbol\tikzperp}}
\newcommand{\goodparallel}{{\mathpalette\goodsymbol\tikzparallel}}
\newcommand{\lvec}[1]{\reflectbox{$\!\vec{\,\reflectbox{$#1$}}$}}
\newcommand{\lrvec}[1]{\mathrlap{\reflectbox{$\!\vec{\,\reflectbox{$#1$}}$}}\,\,\vec{\!\!#1}}
\renewcommand{\overleftarrow}[1]{\lvec{#1}}
\renewcommand{\overrightarrow}[1]{\vec{#1}}
\renewcommand{\overleftrightarrow}[1]{\lrvec{#1}}
\newcommand{\Dd}[0]{\overleftrightarrow{D}}
\newcommand{\Nf}{N_{\!f}}
\makeatletter
\newif\iftag@here
\newcommand*{\taghere}[1][0pt]
{\ifmeasuring@\else
  \global\tag@heretrue
  \tikz[remember picture,overlay]{\coordinate (taghere) at (0pt,#1);}%
\fi}
\def\place@tag{%
    \iftagsleft@
      \kern-\tagshift@
      \iftag@here
        \global\tag@herefalse
        \tikz[remember picture,overlay]%
          {\path (taghere) -| node[anchor=base]{\rlap{\boxz@}} (0pt,0pt);}%
      \else
        \if1\shift@tag\row@\relax
            \rlap{\vbox{%
                \normalbaselines
                \boxz@
                \vbox to\lineht@{}%
                \raise@tag
            }}%
        \else
            \rlap{\boxz@}%
        \fi
        \kern\displaywidth@
      \fi
    \else
      \kern-\tagshift@
      \iftag@here
        \global\tag@herefalse
        \tikz[remember picture,overlay]%
          {\path  (taghere) -|  node[anchor=base]{\llap{\boxz@}} (0pt,0pt);}%
      \else
        \if1\shift@tag\row@\relax
            \llap{\vtop{%
                \raise@tag
                \normalbaselines
                \setbox\@ne\null
                \dp\@ne\lineht@
                \box\@ne
                \boxz@
            }}%
        \else \llap{\boxz@}%
        \fi
      \fi
    \fi
}
\makeatother
\begin{document}
\title{The $\rho$-meson light-cone distribution amplitudes from lattice QCD}
\preprint{{\bf LTH 1108}}
\author[a]{Vladimir~M.~Braun,}
\author[a]{Peter~C.~Bruns,}
\author[a]{Sara~Collins,}
\author[b]{John~A.~Gracey,}
\author[a]{Michael~Gruber,}
\author[a]{Meinulf~G{\"o}ckeler,}
\author[a]{Fabian~Hutzler,}
\author[a]{Paula~P\'erez-Rubio,}
\author[a]{Andreas~Sch{\"a}fer,}
\author[a]{Wolfgang~S{\"o}ldner,}
\author[c]{Andr\'e~Sternbeck}
\author[a]{and Philipp~Wein}
\emailAdd{vladimir.braun@physik.uni-regensburg.de}
\emailAdd{meinulf.goeckeler@physik.uni-regensburg.de}
\emailAdd{fabian.hutzler@physik.uni-regensburg.de}
\emailAdd{andreas.schaefer@physik.uni-regensburg.de}
\affiliation[a]{Institut f{\"u}r Theoretische Physik, Universit{\"a}t Regensburg,\\
Universit{\"a}tsstra{\ss}e 31, 93040 Regensburg, Germany}
\affiliation[b]{Theoretical Physics Division, Department of Mathematical Sciences, University of Liverpool, \\ P.O. Box 147,
Liverpool, L69 3BX, United Kingdom}
\affiliation[c]{Theoretisch-Physikalisches Institut, Friedrich-Schiller-Universit{\"a}t Jena,\\
Max-Wien-Platz 1, 07743 Jena, Germany}
\abstract{We present the results of a lattice study of the normalization 
constants and second moments of the light-cone distribution amplitudes 
of longitudinally and transversely polarized $\rho$ mesons.
The calculation is performed using two flavors of dynamical 
clover fermions 
at lattice spacings between $\unit{0.060}{\femto\meter}$ and 
$\unit{0.081}{\femto\meter}$, different lattice volumes up to 
$m_\pi L = 6.7$ and pion 
masses down to $m_\pi=\unit{150}{\mega\electronvolt}$. 
Bare lattice results are renormalized non-perturbatively using a variant 
of the RI${}^\prime$-MOM scheme and converted to the $\MSbar$ scheme.
The necessary conversion coefficients, which are not available in the
literature, are calculated. 
The chiral extrapolation for the relevant decay constants is 
worked out in detail. We obtain for the ratio of the tensor and vector 
coupling constants $f_\rho^T/f_\rho^{\vphantom{T}} = 0.629(8)$
and the values of the second Gegenbauer moments 
$a_2^\goodparallel = 0.132(27)$ and $a_2^\goodperp = 0.101(22)$ at the scale 
$\mu = \unit{2}{\giga\electronvolt}$ for the 
longitudinally and transversely polarized $\rho$ mesons, respectively. 
The errors include the statistical uncertainty and estimates
of the systematics arising from renormalization. Discretization errors 
cannot be estimated reliably and are not included. In this calculation 
the possibility of $\rho\to\pi\pi$ decay at the smaller pion masses
is not taken into account. 
}

\maketitle%
\flushbottom%
\section{Introduction}%
In recent years exclusive reactions with a large momentum transfer 
to a light vector meson $V = \rho, K^\ast, \phi$ 
in the final state are attracting increasing attention. 
Prominent examples are provided by $B$-meson weak decays, 
$B\to V \pi$, $B\to V \ell \nu_\ell$, $B\to V\gamma$, $B\to V \mu^+\mu^-$, etc. 
Their study constitutes a considerable part of the experimental program 
of the LHCb collaboration at CERN~\cite{Bediaga:2012py}
and the future Belle\,II experiment at the upgraded KEK 
facility~\cite{Aushev:2010bq}. Among these processes, the decays 
$B\to K^\ast\mu^+\mu^-$ and $B_s \to \phi \mu^+\mu^-$ are of particular 
relevance as the angular distributions of the decay products give 
access to a host of observables that are sensitive to new physics, 
see, e.g., ref.~\cite{Blake:2016olu} for a recent review. 
Another example is deeply-virtual exclusive $\rho$-meson production (DVMP) 
in electron-nucleon collisions at high energy,
$e N \to e \rho N$,  that, besides deeply-virtual Compton scattering (DVCS), 
allows one to resolve the transverse distribution of partons inside the 
nucleon. The corresponding cross sections were measured by the HERA 
collider experiments H1 and ZEUS and the fixed target experiments 
HERMES (DESY), CLAS (JLAB), and Hall A (JLAB) at small and moderate 
values of the Bjorken momentum fraction $x_{\mathrm {Bj}}$,
respectively. In the future, exclusive vector meson production will be 
studied with unprecedented precision at the  electron-ion collider 
(EIC)~\cite{Kumericki:2011zc}.

The standard framework for the theoretical description of such processes is based on collinear factorization. In this approach 
the vector mesons are described in terms of light-cone distribution amplitudes (DAs) that specify the distribution of the longitudinal momentum amongst
the quark and antiquark in the valence component of the wave function; 
the transverse degrees of freedom are integrated out. In general, 
meson and baryon DAs are scale-dependent non-perturbative functions and 
their moments (weighted integrals over the 
momentum fractions) are given by matrix elements of local operators. 
From the phenomenological point of view the normalization (representing the 
value of the wave function at the origin) and the first non-trivial Gegenbauer
moment that characterizes the width of the DA are the most relevant quantities. 
For example, knowledge of the second moment of the DA of the longitudinally polarized $\rho$ meson is crucial for global fits
of generalized parton distributions from the DVMP and DVCS data~\cite{Mueller:2013caa}.

The $\rho$-meson coupling to the vector current is known experimentally
and the other parameters were estimated in the past using QCD sum 
rules~\cite{Ball:1996tb}, see also ref.~\cite{Ball:2006nr} for an update. 
Lattice calculations of the tensor coupling have been reported 
in refs.~\cite{Capitani:1999zd,Becirevic:2003pn,Braun:2003jg,Gockeler:2005mh,Allton:2008pn}
and the second moments in ref.~\cite{Arthur:2010xf}. 

In this work we present new results using two flavors of dynamical clover 
fermions at lattice spacings between $\unit{0.060}{\femto\meter}$ and 
$\unit{0.081}{\femto\meter}$, different lattice volumes and pion 
masses down to $m_\pi=\unit{150}{\mega\electronvolt}$. Our approach is 
similar to the strategy used in our paper on the pion DA~\cite{Braun:2015axa}.
In addition to a much larger set of lattices as compared to the previous
studies, a new element of our analysis is a consistent use of non-perturbative
renormalization including mixing with the operators containing total 
derivatives. As the coefficients for the conversion between our 
non-perturbative renormalization scheme on the lattice and the $\MSbar$ scheme 
are not available in the literature for tensor operators, we have 
performed the necessary calculations in continuum perturbation theory
to two loop accuracy. The chiral extrapolation for the relevant 
quantities is worked out in detail.

Although our calculation presents a considerable improvement as compared 
to earlier studies, there are still several issues that we do not address 
in this work. First and foremost, we only consider $\rho$ mesons and leave the 
effects of the $SU(3)$ flavor breaking for a future study. Likewise, we 
do not consider $\omega$ mesons that would require the calculation of 
disconnected diagrams and different techniques. We also do not attempt 
to take into account effects due to the $\rho\to\pi\pi$ decay that 
becomes possible at the smaller pion masses used in our simulations, 
although for the short-distance observables considered in this work
it seems unlikely that such effects are of principal importance.
Last but not least, discretization errors cannot 
be estimated reliably using the set of lattices at our disposal, which 
may be an important problem in such calculations. We 
expect to be able to improve on some of these issues using 
new $N_f=2+1$ lattice configurations that are being generated in the 
framework of the CLS initiative~\cite{Bruno:2014jqa}. This work,
aiming in the long run at smaller lattice spacings with the help of 
open boundary conditions, is in progress. 
 
The presentation is organized as follows. Section~2 is introductory, we 
collect the necessary definitions and specify the quantities that will
be considered in this work. Section~3 contains a list of the correlation 
functions that we compute on the lattice. The lattice ensembles at our 
disposal and the procedure used to extract the signal are described in 
section~4. A non-perturbative calculation of the necessary renormalization 
factors is described in section~5, supplemented by 
appendix~A, where we consider the renormalization of the same operators in 
the continuum and sketch a two loop calculation of the corresponding 
conversion factors. Complete results needed for the evaluation of 
the matching coefficients between our RI${}^\prime$-SMOM scheme 
(defined as in ref.~\cite{Braun:2015axa}) and the $\MSbar$ scheme are 
presented in the auxiliary file attached to the electronic version 
of this paper. Section~6 is devoted to the data analysis and the 
extrapolation to the physical pion mass using, where available, chiral 
effective field theory expressions derived in appendix~B. The final 
section~7 contains a summary of our results and a discussion. 

\section{General formalism}
\subsection{Continuum formulation}
The $\rho$ meson has two independent leading twist (twist two) DAs, 
$\phi^\goodparallel_\rho$ and 
$\phi^\goodperp_\rho$~\cite{Chernyak:1983ej}, 
corresponding to longitudinal and transverse polarization, respectively. 
Neglecting isospin breaking and electromagnetic effects, the DAs of 
charged $\rho^\pm$ and neutral $\rho^0$ mesons are related 
so that it is sufficient to consider one of them, for example, $\rho^+$.
The DAs are defined as meson-to-vacuum matrix elements of renormalized 
non-local quark-antiquark light-ray operators,
\begin{subequations}\label{def:phi}%
\begin{align}%
 \langle0| \bar{d}(z_1n) \slashed{n} [z_1n, z_2n] u(z_2&n) 
    |\rho^+(p,\lambda)\rangle 
\nonumber \\
&= 
 m_\rho f_\rho (e^{(\lambda)} \cdot n) \int_0^1 \!dx\, 
e^{-i p \cdot n (z_1(1-x)+z_2x)} \phi^\goodparallel_{\rho}(x,\mu)\,,
\\
 e_{\goodperp,\mu}^{(\lambda^\prime)}n_\nu 
\langle 0| \bar{d}(z_1n) \sigma^{\mu\nu} [z_1n, & z_2n] u(z_2n) 
    |\rho^+(p,\lambda)\rangle 
\nonumber \\
&=
if_\rho^T \bigl(e_\goodperp^{(\lambda^\prime)} \cdot e_\goodperp^{(\lambda)}\bigr)(p \cdot n)
 \int_0^1 \!dx\, e^{-i p \cdot n (z_1(1-x)+z_2x)}\phi^\goodperp_{\rho}(x,\mu)\,,
\end{align}%
\end{subequations}%
where $\sigma_{\mu\nu} = \frac{i}{2}[\gamma_\mu,\gamma_\nu]$,  $z_{1,2}$ 
are real numbers, $n^\mu$ is an auxiliary light-like vector ($n^2=0$), 
and $|\rho^+(p,\lambda)\rangle$ is the state of the $\rho^+$ meson 
with on-shell momentum $p^2=m_\rho^2$ and polarization $\lambda$. 
The straight-line path-ordered Wilson line connecting the quark 
fields, $[z_1n,z_2n]$, is inserted to ensure gauge invariance.
The $\rho$-meson polarization vector $e^{(\lambda)}_\mu$ has the 
following properties:
\begin{equation}%
 e^{(\lambda)} \cdot p = 0 \,,   \quad
 \sum_\lambda e^{(\lambda)}_\mu  e^{(\lambda)*}_\nu = 
  - \eta_{\mu\nu} + \frac{p_\mu p_\nu}{m_\rho^2}\,,
\end{equation}
and we use the notation
\begin{equation}
 e_{\goodperp,\mu}^{(\lambda)} = e_\mu^{(\lambda)} - 
\frac{e^{(\lambda)} \cdot n}{p \cdot n}
\biggl(p_\mu-\frac{m_\rho^2}{p \cdot n}n_\mu\biggr)\,.
\end{equation}%
The variable $x$ has the meaning of the fraction of the 
$\rho$ meson's light-cone momentum $p \cdot n$ which is carried 
by the $u$-quark, 
whereas $1-x$ is the momentum fraction carried by the antiquark $\bar{d}$, 
and $\mu$ is the renormalization scale 
(we assume the $\MSbar$ scheme). 
The scale dependence will often be suppressed in what follows. 

The couplings $f_\rho^{\vphantom{T}}$ and $f_\rho^T$ appearing in 
\eqref{def:phi} are defined as matrix elements of local operators:
\begin{subequations}%
\begin{align}%
 \langle0| \bar{d}(0) \gamma_{\mu} u(0) |\rho^+(p,\lambda)\rangle &= f_\rho m_\rho e^{(\lambda)}_\mu\,,
\\
 \langle0| \bar{d}(0) \sigma_{\mu\nu} u(0) |\rho^+(p,\lambda)\rangle &= 
if^T_\rho \bigl(e^{(\lambda)}_\mu p^{\vphantom(}_\nu - e^{(\lambda)}_\nu p^{\vphantom(}_\mu\bigr)\,.
\end{align}%
\end{subequations}%
In the following, we will refer to them as vector and tensor couplings, 
respectively. The vector coupling $f_\rho$ is scale independent and 
can be extracted from experiment, see appendix~C in ref.~\cite{Straub:2015ica} 
for a detailed discussion. One obtains~\cite{Straub:2015ica}
\begin{align} \label{eq:experiment}
f_{\rho^+} = \unit{(210 \pm 4)}{\mega\electronvolt} \,,
&& f^{(u)}_{\rho^0} = \unit{(221.5\pm 3)}{\mega\electronvolt} \,,
&& f^{(d)}_{\rho^0} = \unit{(209.7\pm 3)}{\mega\electronvolt} \,,
\end{align}
where for the neutral $\rho$ meson we quote separate values for the 
$\bar u u $ and $\bar d d$ currents.  The difference in the 
given three values is due to isospin breaking and electromagnetic 
corrections, which will be neglected throughout this study.

The tensor coupling $f_\rho^T$ is scale dependent and is not directly 
accessible from experiment. To leading order one obtains
\begin{align}%
 \label{eq:fTscaling}
 f^T_\rho(\mu) &= f^T_\rho(\mu_0) \biggl(\frac{\alpha_s(\mu)}{\alpha_s(\mu_0)}\biggr)^{C_F/\beta_0}\,,
\end{align}%
where $C_F=(N_c^2-1)/(2N_c)$, $\beta_0=(11N_c-2N_f)/3$,
$N_c=3$ is the number of colors and $N_f$ the number of active flavors.

The DAs are normalized to unity, 
\begin{align}%
 \int_0^1 \!dx\, \phi^{\goodparallel,\goodperp}_\rho(x) &= 1\,,
\label{phi:norm}
\end{align}%
and, neglecting isospin breaking effects, are symmetric under the 
interchange of the momentum fractions of the quark and the antiquark,
\begin{align}%
 \phi^{\goodparallel,\goodperp}_\rho(x) &=  
\phi^{\goodparallel,\goodperp}_\rho(1-x)\,.
\label{phi:x->1-x}
\end{align}%
For convenience we introduce a generic notation 
$\langle \cdots \rangle^{\goodparallel,\goodperp}$
for the moments of the DAs defined as weighted integrals of the type
\begin{align}
\langle x^k (1-x)^l \rangle^{\goodparallel,\goodperp} 
& =  \int_0^1 \!dx\, x^k (1-x)^l\, \phi^{\goodparallel,\goodperp}_\rho(x)\,.   
\label{phi:moments}
\end{align}
The symmetry property \eqref{phi:x->1-x} implies
\begin{align}
\langle (1-x)^k x^l \rangle^{\goodparallel , \goodperp} = 
   \langle (1-x)^l x^k \rangle^{\goodparallel , \goodperp}\,,
\end{align}
and in addition we have the (momentum conservation) constraint
\begin{align}
\langle (1-x)^k x^l \rangle^{\goodparallel , \goodperp} = 
\langle (1-x)^{k+1} x^l \rangle^{\goodparallel , \goodperp} +
\langle (1-x)^k x^{l+1} \rangle^{\goodparallel , \goodperp}\,.
\end{align}
Hence the set of moments \eqref{phi:moments} for positive integers 
$k,l$ is overcomplete. We introduce the variable
\begin{align}
 \xi = 2x-1 \,,
\end{align}
corresponding to the \emph{difference} of the momentum fraction between 
the quark and the antiquark and consider $\xi$ moments
\begin{align}
\langle \xi^n \rangle^{\goodparallel , \goodperp} =  
\langle (2x-1)^n \rangle^{\goodparallel , \goodperp}\,,\qquad n=2,4,6,\ldots\,,
\end{align}
or, alternatively, Gegenbauer moments 
\begin{align}
a_n^{\goodparallel , \goodperp} = \frac{2(2n+3)}{3(n+1)(n+2)}
\langle\, C_n^{3/2}(2x-1)\, \rangle^{\goodparallel , \goodperp}
\end{align}
as independent non-perturbative parameters. The two sets are related by  
simple algebraic relations, e.g.,
\begin{align}%
\label{eq:a2xi2}
 a_2^{\goodparallel,\goodperp} &= \frac{7}{12}\bigl(5\langle\xi^2\rangle^{\goodparallel,\goodperp}-1\bigr)\,.
\end{align}%
The rationale for using Gegenbauer moments is that they have autonomous scale dependence at the one loop level 
\begin{subequations}
\begin{align}%
 a^\goodparallel_n(\mu) &= a^\goodparallel_n(\mu_0) \biggl(\frac{\alpha_s(\mu)}{\alpha_s(\mu_0)}\biggr)^{\gamma^{\goodparallel(0)}_n/(2\beta_0)}\,,
\\
 a^\goodperp_n(\mu)     &= a^\goodperp_n(\mu_0)     \biggl(\frac{\alpha_s(\mu)}{\alpha_s(\mu_0)}\biggr)^{\bigl(\gamma^{\goodperp(0)}_n-\gamma^{\goodperp(0)}_0\bigr)/(2\beta_0)}\,.
\end{align}
\end{subequations}
The anomalous dimensions are given by
\begin{align}%
 \gamma^{\goodparallel(0)}_n = 8C_F \biggl( \sum_{k=1}^{n+1} \frac{1}{k} - \frac{3}{4} - \frac{1}{2(n+1)(n+2)} \biggr)\,,
&&
 \gamma^{\goodperp(0)}_n     = 8C_F \biggl( \sum_{k=1}^{n+1} \frac{1}{k} - \frac{3}{4} \biggr)\,.
\end{align}%
As Gegenbauer polynomials form a complete set of functions, the DAs can be 
written as an expansion 
\begin{align}%
 \phi_\rho^{\goodparallel,\goodperp}(x,\mu) &= 6x(1-x) 
\biggl[ 1 + \smashoperator{\sum_{n=2,4,\ldots}^\infty} a^{\goodparallel,\goodperp}_n(\mu) C_n^{3/2}(2x-1)\biggr]\,.
\end{align}%
Typical integrals that one encounters in applications can also be 
expressed in terms of the Gegenbauer coefficients, e.g., 
\begin{align}
\int_0^1\!\frac{dx}{1-x}\phi^{\goodparallel,\goodperp}_\rho(x,\mu)&=  
3\, \biggl[1 + \smashoperator{\sum_{n=2,4,\ldots}^\infty} a^{\goodparallel,\goodperp}_n(\mu)\biggr]\,.
\end{align}
Since the anomalous dimensions increase with $n$, the higher-order 
contributions in the Gegenbauer expansion are suppressed at large 
scales so that asymptotically only the leading term survives, 
usually referred to as the asymptotic DA:
\begin{align}%
 \label{def:phias}
 \phi^{\goodparallel,\goodperp}_\rho(x,\mu\to\infty) &= 
\phi_{\text{as}}(x) = 6x(1-x)\,.
\end{align}
Beyond the leading order, higher Gegenbauer coefficients $a_n$ 
mix with the lower ones, $a_k,\,  k <n$~\cite{Mikhailov:1984ii,Mueller:1993hg}. 
This implies, in particular, that Gegenbauer coefficients with higher 
values of $n$ are generated by the evolution even if they vanish
at a low reference scale. This effect is numerically small, however, 
so that it is usually reasonable to employ the Gegenbauer 
expansion to some fixed order.

\subsection{Lattice formulation}%

From now on we work in Euclidean space, using the same conventions
as in ref.~\cite{Braun:2015axa}. 
The renormalized light-ray operators entering the definition of the DAs 
are defined as the generating functions for the corresponding 
renormalized local operators, cf.\ ref.~\cite{Balitsky:1987bk}. This means 
that moments of the DAs, by construction, are given by matrix elements 
of local operators and can be evaluated on the lattice using the 
Euclidean version of QCD.

Our aim in this work is to calculate the couplings 
$f_\rho^{\vphantom{T}}$, $f_\rho^T$ and the second DA moments. 
To this end we define bare operators
\begin{subequations} \label{def:currents}
\begin{align}%
V_\mu (x) &= \bar{d}(x) \gamma_\mu u(x) \,,
\\
T_{\mu \nu} (x) &= \bar{d}(x) \sigma_{\mu \nu} u(x) 
\end{align}%
\end{subequations}%
and
\begin{subequations}\label{def:bareops}%
\begin{align}%
V^\pm_{\mu \nu \rho} (x) &= \bar{d}(x) \gamma_\mu \left( 
\overleftarrow{D}_\nu \overleftarrow{D}_\rho + 
\overrightarrow{D}_\nu \overrightarrow{D}_\rho \pm 
2 \overleftarrow{D}_\nu \overrightarrow{D}_\rho \right) u(x) \,,
\label{def:bareops_long} \\
T^\pm_{\mu \nu \rho \sigma} (x) &= \bar{d}(x) \sigma_{\mu \nu} \left(
\overleftarrow{D}_\rho \overleftarrow{D}_\sigma + 
\overrightarrow{D}_\rho \overrightarrow{D}_\sigma \pm 
2 \overleftarrow{D}_\rho \overrightarrow{D}_\sigma \right) u(x) \,.
\label{def:bareops_trans}
\end{align}%
\end{subequations}%
On the lattice the covariant derivatives will be replaced by their
discretized versions.

Projection onto the leading twist corresponds to symmetrization 
over the maximal possible set of Lorentz indices and subtraction 
of traces. The operation of symmetrization and trace subtraction 
will be indicated by enclosing the involved Lorentz indices in 
parentheses, for instance,
$O_{(\mu \nu)} = \tfrac{1}{2}(O_{\mu \nu} + O_{\nu \mu}) 
- \tfrac{1}{4} \delta_{\mu \nu} O_{\lambda \lambda}$.
Note that for the operators involving the $\sigma_{\mu \nu}$-matrix  
also those traces have to be subtracted which correspond to index pairs
where one of the indices equals $\mu$ or $\nu$.

Using the shorthand 
$\Dd_{\mu} = \overrightarrow{D}_\mu - \overleftarrow{D}_\mu$, 
the operator $V^-_{(\mu \nu \rho)}(x)$ can be rewritten as
\begin{equation}%
V^-_{(\mu \nu \rho)}(x) = \bar{d}(x) \gamma_{(\mu} 
\Dd_{\vphantom{(}\nu} \Dd_{\rho)} u(x) 
\end{equation}%
and its matrix element between the vacuum and the $\rho$ state is proportional
to the bare value of the second moment $\langle \xi^2 \rangle^{\goodparallel}$:
\begin{align}
\langle 0|V^-_{(\mu \nu \rho)}  |\rho^+(p,\lambda)\rangle &= \mathcal{N}_{(\mu \nu \rho)} \langle \xi^2 \rangle^{\goodparallel}_{\text{bare}} \,,
\end{align} 
where $\mathcal{N}_{(\mu \nu \rho)}$ is a kinematical prefactor. 
The operator $V^+_{(\mu \nu \rho)}(x)$ in the continuum reduces 
to the second derivative of the vector current,
\begin{equation}\label{eq:totald}%
V^+_{(\mu \nu \rho)}(x) = 
\partial_{(\mu} \partial_{\vphantom{(}\nu} \bar{d} (x) \gamma_{\rho)} u(x) \,,
\end{equation}%
so that 
\begin{align}
\langle 0|V^+_{(\mu \nu \rho)}  |\rho^+(p,\lambda)\rangle  & =  \mathcal{N}_{(\mu \nu \rho)}\langle 1^2 \rangle^{\goodparallel}_{\text{bare}}
\end{align} 
with the same prefactor. While in the  continuum 
$\langle 1^2 \rangle^{\goodparallel}_{\text{bare}} =1$ by construction, 
this is no longer true on the lattice because the Leibniz rule holds 
for discretized derivatives only up to lattice artefacts and hence 
\eqref{eq:totald} is violated.
As we will see below, the deviation from unity for the \emph{renormalized}
$\langle 1^2 \rangle^{\goodparallel}$ is small. Nevertheless, it still 
has to be taken into account and affects the relation between 
$\langle \xi^2 \rangle^{\goodparallel}$ and the Gegenbauer moment at 
finite lattice spacing~\cite{Braun:2015axa}:
\begin{align}
 a_2^{\goodparallel} &= \frac{7}{12}\bigl(5\langle\xi^2\rangle^{\goodparallel} -\langle 1^2 \rangle^{\goodparallel}\bigr)\,.
\end{align}
The situation with the tensor operators $T^\pm_{\mu (\nu \rho \sigma)}$ 
and the corresponding matrix elements
$\langle \cdots \rangle^{\goodperp}$ is similar.

The operators $V^-_{(\mu \nu \rho)}$ and $V^+_{(\mu \nu \rho)}$ mix 
under renormalization even in the continuum, as do 
$T^-_{\mu (\nu \rho \sigma)}$ and $T^+_{\mu (\nu \rho \sigma)}$.
Additional mixing could result from the fact that the continuous $O(4)$
symmetry of Euclidean space is reduced to the discrete $H(4)$ symmetry of
the hypercubic lattice. This is particularly worrisome if operators of 
lower dimension are involved. Fortunately, in the case at hand it is 
possible to avoid additional mixing by using suitably chosen operators,
which will be detailed below.

\section{Lattice correlation functions}%

The basic objects from which moments of the $\rho$ DAs can be extracted 
on the lattice are 2-point correlation functions. In order to ``create'' 
the $\rho$ meson we use the interpolating current $\mathscr{V}_\nu (x)$,
which is defined as $V_\nu (x)$ with smeared quark fields. For details
of our smearing algorithm see section~\ref{sec:details}. Let $\cO$ be 
a local (unsmeared) operator, e.g., one of the operators defined in 
eq.~\eqref{def:bareops} above. One then obtains for the 2-point function 
in the region where the ground state dominates%
\begin{equation} \label{eq:twopoint}
 a^3 \sum_\mathbf{x} e^{-i\mathbf{p}\cdot\mathbf{x}} \langle \mathcal{O}(t,\mathbf{x}) 
 \mathscr{V}^\dagger_\nu(0) \rangle = \frac{1}{2E}
 A(\cO,\mathscr{V}_\nu\mid\mathbf{p}) \bigl(e^{-Et} + 
  \sigma \, \sigma_\cO \, (-1)^{n_t} e^{-E(T-t)}\bigr)
\end{equation}%
with%
\begin{align} \label{eq:amplitude}
 A(\cO,\mathscr{V}_\nu\mid\mathbf{p}) = 
\sum_\lambda \langle0| \cO(0) |\rho^+(\mathbf{p},\lambda)\rangle 
\langle\rho^+(\mathbf{p},\lambda)| \mathscr{V}_\nu^\dagger(0) 
|0\rangle\,.
\end{align}%
Here $T$ is the time extent of the lattice, $a$ is the lattice spacing,
and $E$ denotes the energy of the $\rho$ state. The sign factors $\sigma$ 
are determined by the Dirac matrices in the creation operator 
(which is in our case always $\gamma_\nu$), while $\sigma_\cO$ are the 
analogous factors for $\cO$ (see table~\ref{signs.dirac}), and $n_t$ 
is the number of time derivatives in $\cO$.%
\begin{table}[t]%
\centering%
\begin{tabular}{ccccccccc}%
\toprule
$\Gamma$ &  $\mathds{1}$  & $\gamma_j$ & $\gamma_4$ & $\gamma_j \gamma_5$ & 
$\gamma_4 \gamma_5$ & $\gamma_5$ & $\gamma_j \gamma_k$ & $\gamma_j \gamma_4$ \\
\midrule
$\sigma$ &  $1$ & $1$ & $-1$ & $-1$ & $1$ & $-1$ & $1$ & $-1$ \\
\bottomrule
\end{tabular}%
\caption{\label{signs.dirac}Sign factors $\sigma$ for the different Dirac matrices
$\Gamma$. Here $j,k \in \{ 1,2,3 \}$ and $j \neq k$.}%
\end{table}%

For the decay constants and the second DA moments of the $\rho$ meson we 
have to evaluate the following set of correlation functions:%
\begin{subequations}%
\begin{align}%
 \mathscr{C}_{\mu_1\nu}(t,\mathbf{p}) &= a^3 \sum_\mathbf{x} 
 e^{-i\mathbf{p} \cdot \mathbf{x}} \langle \mathscr{V}_{\mu_1}(t,\mathbf{x}) 
        \mathscr{V}^\dagger_\nu(0) \rangle\,,\label{eq:Operators-11}\\
C_{\mu_1\nu}(t,\mathbf{p}) &= a^3 \sum_\mathbf{x} 
 e^{-i\mathbf{p} \cdot \mathbf{x}} \langle V_{\mu_1}(t,\mathbf{x}) 
     \mathscr{V}^\dagger_\nu(0) \rangle\,,\label{eq:Operators-12}\\
C_{\mu_0\mu_1\nu}(t,\mathbf{p}) &= a^3 \sum_\mathbf{x} 
 e^{-i\mathbf{p} \cdot \mathbf{x}} \langle T_{\mu_0\mu_1}(t,\mathbf{x}) 
     \mathscr{V}^\dagger_\nu(0) \rangle\,,\label{eq:Operators-13}\\
C_{\mu_1\mu_2\mu_3\nu}^\pm(t,\mathbf{p}) &= a^3 \sum_\mathbf{x} 
 e^{-i\mathbf{p} \cdot \mathbf{x}} \langle V^\pm_{\mu_1\mu_2\mu_3}(t,\mathbf{x}) 
       \mathscr{V}^\dagger_\nu(0) \rangle\,,\label{eq:Operators-14}\\
C_{\mu_0\mu_1\mu_2\mu_3\nu}^\pm(t,\mathbf{p}) &= a^3 \sum_\mathbf{x} 
 e^{-i\mathbf{p} \cdot \mathbf{x}} \langle T^\pm_{\mu_0\mu_1\mu_2\mu_3}(t,\mathbf{x}) 
      \mathscr{V}^\dagger_\nu(0) \rangle\,.\label{eq:Operators-15}
\end{align}%
\end{subequations}%
\subsection{Decay constants}%
In order to determine the leading twist $\rho$-meson couplings we 
use the correlation functions 
\begin{subequations}%
\begin{align}%
 \mathscr{C}_{jj}(t,\mathbf{0}) &= Z_\rho \frac{1}{2m_\rho} 
\bigl(e^{-m_\rho t}+e^{-m_\rho(T-t)}\bigr) + \cdots \,,\\
 C_{jj}(t,\mathbf{0}) &= m_\rho f_\rho \sqrt{Z_\rho} \frac{1}{2m_\rho} 
\bigl(e^{-m_\rho t}+e^{-m_\rho(T-t)}\bigr) + \cdots \,,\\
 C_{4jj}(t,\mathbf{0}) &= - i m_\rho f^T_\rho \sqrt{Z_\rho} \frac{1}{2m_\rho} 
\bigl(e^{-m_\rho t}-e^{-m_\rho(T-t)}\bigr) 
+ \cdots \,,
\end{align}%
\end{subequations}
with $j=1,2,3$, assuming the dominance of the lowest one-particle state.

In the actual fits we average over the forward and backward running states.
As in our simulations the signal disappears in the noise well before the
middle point $t=T/2$ in the time direction is reached (see 
figure~\ref{fig_range} for an example), the ``mixing'' of these two 
contributions is completely negligible. Therefore we work with simple
exponential fits,
\begin{subequations}%
\begin{align}%
\frac13 \sum_{j=1}^3 \hat{t}_+ \mathscr{C}_{jj}(t,\mathbf{0}) 
  &= \frac{Z_\rho}{2m_\rho} e^{-m_\rho t}\,,\label{eq:corrZ}\\
\frac13 \sum_{j=1}^3 \hat{t}_+ C_{jj}(t,\mathbf{0}) 
  &= f_\rho \frac{\sqrt{Z_\rho}}{2} e^{-m_\rho t}\,,\label{eq:corrf}\\
 \frac13 \sum_{j=1}^3 \hat{t}_-C_{4jj}(t,\mathbf{0}) 
  &= - i f^T_\rho \frac{\sqrt{Z_\rho}}{2} e^{-m_\rho t}\,, \label{eq:corrfT}
\end{align}%
\end{subequations}%
where the averaging operator $\hat{t}_\pm$ is defined as
\begin{equation}
\hat{t}_\pm C(t,\mathbf{p}) = 
  \tfrac12 \bigl( C(t,\mathbf{p}) \pm C(T-t,\mathbf{p}) \bigr)\,.
\end{equation}

The decay constants $f_\rho^{\vphantom{T}}$ and $f^T_\rho$ 
can be obtained by simultaneously fitting the correlation functions 
\eqref{eq:corrZ}--\eqref{eq:corrfT}. The result for the mass
is then dominated by the two correlation functions 
\eqref{eq:corrf} and \eqref{eq:corrfT} that contain an unsmeared operator 
at the sink, because they have much smaller statistical errors. 
However, they exhibit larger contributions from excited states so that the 
isolation of the ground state is less reliable. Therefore we first
fit the correlator with a smeared operator at the sink, \eqref{eq:corrZ}, 
to extract $Z_\rho$ and $m_\rho$. These values are then inserted in 
eqs.~\eqref{eq:corrf} and \eqref{eq:corrfT} in order to obtain 
$f_\rho^{\vphantom{T}}$ and $f^T_\rho$ as well as 
$f_\rho^T/f_\rho^{\vphantom{T}}$ from a second fit. This procedure is 
repeated on every bootstrap sample allowing an estimation of the 
statistical error.
\subsection{Second moments --- the longitudinal case}%
Multiplets of twist-2 operators suitable for the evaluation of the
second longitudinal moments consist of the operators 
\begin{subequations}\label{opslong}%
\begin{align}%
\mathcal{O}_{1}^\pm &= V^\pm_{\{234\}}\,,\\
\mathcal{O}_{2}^\pm &= V^\pm_{\{134\}}\,,\\
\mathcal{O}_{3}^\pm &= V^\pm_{\{124\}}\,,\\
\mathcal{O}_{4}^\pm &= V^\pm_{\{123\}}\,.
\end{align}%
\end{subequations}%
Here and in the following $\{ \cdots \}$ denotes symmetrization of the 
enclosed $n$ indices with an overall factor $1/n!$ included.
The two multiplets $\mathcal{O}_1^+ , \ldots , \mathcal{O}_4^+$ and
$\mathcal{O}_1^- , \ldots , \mathcal{O}_4^-$ both
transform according to the irreducible representation 
$\tau^{(4)}_2$ of the hypercubic group $H(4)$~\cite{Gockeler:1996mu}. 
Their symmetry properties ensure that under renormalization they can 
only mix with each other, but mixing with 
additional operators of the same or lower dimension is forbidden. 
The amplitudes \eqref{eq:amplitude} of the 2-point functions 
\eqref{eq:twopoint} where $\mathcal O$ is one member of these multiplets 
are related to the amplitudes where $\mathcal O$ is a component of the
vector current $V_\mu$ by
\begin{subequations} \label{eq:amplong}
\begin{align}%
A(\cO_1^\pm,\mathscr{V}_\nu\mid\mathbf{p})
&= - \tfrac{1}{3} R^\goodparallel_{\pm} \bigl(
p_2 p_3 \, A( V_4,\mathscr{V}_\nu\mid\mathbf{p})+ 
i p_2 E \, A( V_3,\mathscr{V}_\nu\mid\mathbf{p})+ 
i p_3 E \, A( V_2,\mathscr{V}_\nu\mid\mathbf{p})\bigr) \,,
\\
A(\cO_2^\pm,\mathscr{V}_\nu\mid\mathbf{p})
&= - \tfrac{1}{3} R^\goodparallel_{\pm} \bigl(
p_1 p_3 \, A( V_4,\mathscr{V}_\nu\mid\mathbf{p})+ 
i p_1 E \, A( V_3,\mathscr{V}_\nu\mid\mathbf{p})+ 
i p_3 E \, A( V_1,\mathscr{V}_\nu\mid\mathbf{p})\bigr) \,,
\\
A(\cO_3^\pm,\mathscr{V}_\nu\mid\mathbf{p})
&= - \tfrac{1}{3} R^\goodparallel_{\pm} \bigl(
p_1 p_2 \, A( V_4,\mathscr{V}_\nu\mid\mathbf{p})+ 
i p_2 E \, A( V_1,\mathscr{V}_\nu\mid\mathbf{p})+ 
i p_1 E \, A( V_2,\mathscr{V}_\nu\mid\mathbf{p})\bigr) \,,
\\
A(\cO_4^\pm,\mathscr{V}_\nu\mid\mathbf{p})
&= - \tfrac{1}{3} R^\goodparallel_{\pm} \bigl(
p_1 p_2 \, A( V_3,\mathscr{V}_\nu\mid\mathbf{p})+ 
p_1 p_3 \, A( V_2,\mathscr{V}_\nu\mid\mathbf{p})+ 
p_2 p_3 \, A( V_1,\mathscr{V}_\nu\mid\mathbf{p})\bigr) \,.
\end{align}%
\end{subequations}%
In order to be able to write these and some of the following formulae
in a compact form we have introduced the notation $R^\goodparallel_{\pm}$,
where $R^\goodparallel_+$ ($R^\goodparallel_-$) is the bare value of 
$\langle 1^2 \rangle^{\goodparallel}$ 
($\langle \xi^2 \rangle^{\goodparallel}$).

We will try to increase the signal-to-noise ratio by considering only
correlation functions with the smallest non-zero momentum in one 
spatial direction, which is equal to $2 \pi/L$ on a lattice of spatial
extent $L$. Therefore we exclude $\cO_4^\pm$ from our calculation.
After averaging over all suitable combinations as well as over forward
and backward running states, the second longitudinal moments can be 
obtained from the ratio
\begin{equation}\label{eq-long}%
 \frac16\sum_{j=1}^3\smash{\sum_{\substack{\vphantom{j}k=1\\k\neq j}}^3}
 \frac{\hat{p}_-\hat{t}_-C_{\{4jk\}k}^\pm\bigl(t,\frac{2\pi}{L}\mathbf{e}_j\bigr)}{\hat{p}_+\hat{t}_+C_{kk}\bigl(t,\frac{2\pi}{L}\mathbf{e}_j\bigr)}
 = -\frac{2\pi}{L}E\frac13iR^\goodparallel_\pm\,,
\end{equation}%
where momentum averaging is accounted for by the operator $\hat{p}_\pm$:%
\begin{align}%
 \hat{p}_\pm C(t,\mathbf{p}) &= \tfrac12 \bigl( C(t,\mathbf{p}) \pm C(t,-\mathbf{p}) \bigr)\,.
\end{align}%
\subsection{Second moments --- the transverse case}%
In the transverse case we consider the following multiplets: 
\begin{subequations}\label{mp4p9}%
\begin{align}%
 \mathcal{O}_{1,T}^\pm &= T^\pm _{13\{32\}} + T^\pm _{23\{31\}} - T^\pm _{14\{42\}} - T^\pm _{24\{41\}}  \,,\\
 \mathcal{O}_{2,T}^\pm &= T^\pm _{12\{23\}} + T^\pm _{32\{21\}} - T^\pm _{14\{43\}} - T^\pm _{34\{41\}}  \,,\\
 \mathcal{O}_{3,T}^\pm &= T^\pm _{12\{24\}} + T^\pm _{42\{21\}} - T^\pm _{13\{34\}} - T^\pm _{43\{31\}}  \,,\\
 \mathcal{O}_{4,T}^\pm &= T^\pm _{21\{13\}} + T^\pm _{31\{12\}} - T^\pm _{24\{43\}} - T^\pm _{34\{42\}}  \,,\\
 \mathcal{O}_{5,T}^\pm &= T^\pm _{21\{14\}} + T^\pm _{41\{12\}} - T^\pm _{23\{34\}} - T^\pm _{43\{32\}}  \,,\\
 \mathcal{O}_{6,T}^\pm &= T^\pm _{31\{14\}} + T^\pm _{41\{13\}} - T^\pm _{32\{24\}} - T^\pm _{42\{23\}}  \,.    
\end{align}%
\end{subequations}%
The two multiplets $\mathcal{O}_{1,T}^+ , \ldots , \mathcal{O}_{6,T}^+$ and
$\mathcal{O}_{1,T}^- , \ldots , \mathcal{O}_{6,T}^-$ both
transform according to the irreducible representation 
$\tau^{(6)}_2$ of the hypercubic group $H(4)$. As in the case of the
multiplets \eqref{opslong}, mixing with additional operators of the 
same or lower dimension is forbidden by symmetry. 
The amplitudes \eqref{eq:amplitude} of the 2-point functions 
\eqref{eq:twopoint} where $\mathcal O$ is one member of the multiplets
\eqref{mp4p9} are related to the amplitudes where $\mathcal O$ is a 
component of the tensor current $T_{\mu \nu}$ by
\begin{subequations} \label{eq:amptrans}
\begin{align}%
A(\cO_{1,T}^\pm,\mathscr{V}_\nu\mid\mathbf{p})
& = - R^\goodperp_{\pm} \bigl(\!\begin{aligned}[t]&
  p_2 p_3 \, A( T_{13},\mathscr{V}_\nu\mid\mathbf{p})
+ p_1 p_3 \, A( T_{23},\mathscr{V}_\nu\mid\mathbf{p})
\\&
\mathllap{{}+{}} i p_2 E \, A( T_{41},\mathscr{V}_\nu\mid\mathbf{p})
+ i p_1 E \, A( T_{42},\mathscr{V}_\nu\mid\mathbf{p})
\bigr)\,,\taghere\end{aligned}\\
A(\cO_{2,T}^\pm,\mathscr{V}_\nu\mid\mathbf{p})
& = - R^\goodperp_{\pm} \bigl(\!\begin{aligned}[t]&
  p_2 p_3 \, A( T_{12},\mathscr{V}_\nu\mid\mathbf{p})
+ p_1 p_2 \, A( T_{32},\mathscr{V}_\nu\mid\mathbf{p})
\\&
\mathllap{{}+{}} i p_3 E \, A( T_{41},\mathscr{V}_\nu\mid\mathbf{p})
+ i p_1 E \, A( T_{43},\mathscr{V}_\nu\mid\mathbf{p})
\bigr) \,,\taghere\end{aligned}\\
A(\cO_{3,T}^\pm,\mathscr{V}_\nu\mid\mathbf{p})
& = - R^\goodperp_{\pm} \bigl(\!\begin{aligned}[t]&
  p_1 p_2 \, A( T_{42},\mathscr{V}_\nu\mid\mathbf{p})
- p_1 p_3 \, A( T_{43},\mathscr{V}_\nu\mid\mathbf{p})
\\&
\mathllap{{}+{}} i p_2 E \, A( T_{12},\mathscr{V}_\nu\mid\mathbf{p})
+ i p_3 E \, A( T_{31},\mathscr{V}_\nu\mid\mathbf{p})
\bigr) \,,\taghere\end{aligned}\\
A(\cO_{4,T}^\pm,\mathscr{V}_\nu\mid\mathbf{p})
& = - R^\goodperp_{\pm} \bigl(\!\begin{aligned}[t]&
  p_1 p_3 \, A( T_{21},\mathscr{V}_\nu\mid\mathbf{p})
+ p_1 p_2 \, A( T_{31},\mathscr{V}_\nu\mid\mathbf{p})
\\&
\mathllap{{}+{}} i p_3 E \, A( T_{42},\mathscr{V}_\nu\mid\mathbf{p})
+ i p_2 E \, A( T_{43},\mathscr{V}_\nu\mid\mathbf{p})
\bigr) \,,\taghere\end{aligned}\\
A(\cO_{5,T}^\pm,\mathscr{V}_\nu\mid\mathbf{p})
& = - R^\goodperp_{\pm} \bigl(\!\begin{aligned}[t]&
  p_1 p_2 \, A( T_{41},\mathscr{V}_\nu\mid\mathbf{p})
- p_2 p_3 \, A( T_{43},\mathscr{V}_\nu\mid\mathbf{p})
\\&
\mathllap{{}+{}} i p_1 E \, A( T_{21},\mathscr{V}_\nu\mid\mathbf{p})
+ i p_3 E \, A( T_{32},\mathscr{V}_\nu\mid\mathbf{p})
\bigr) \,,\taghere\end{aligned}\\
A(\cO_{6,T}^\pm,\mathscr{V}_\nu\mid\mathbf{p})
& = - R^\goodperp_{\pm} \bigl(\!\begin{aligned}[t]&
  p_1 p_3 \, A( T_{41},\mathscr{V}_\nu\mid\mathbf{p})
- p_2 p_3 \, A( T_{42},\mathscr{V}_\nu\mid\mathbf{p})
\\&
\mathllap{{}+{}} i p_1 E \, A( T_{31},\mathscr{V}_\nu\mid\mathbf{p})
+ i p_2 E \, A( T_{23},\mathscr{V}_\nu\mid\mathbf{p})
\bigr) \,.\taghere\end{aligned}
\end{align}%
\end{subequations}%
Here $R^\goodperp_+$ ($R^\goodperp_-$) is the bare value of 
$\langle 1^2 \rangle^{\goodperp}$ ($\langle \xi^2 \rangle^{\goodperp}$).

As in the longitudinal case, we only consider correlation functions 
with the smallest non-zero momentum in one spatial direction and perform
averages similar to those in eq.~\eqref{eq-long}. This leads to the 
following ratio for the second transverse moments:
\begin{align}\label{eq-trans}%
 \!\begin{aligned}[b]
 \frac16\sum_{j=1}^3\smash{\sum_{\substack{\vphantom{k}l=1\\\vphantom{k}l\neq j}}^3}\smash{\sum_{\substack{\vphantom{l}k=1\\k\neq j\\k\neq l}}^3} \biggl(
 &\frac{\hat{p}_-\hat{t}_+\bigl(C^\pm_{4\{4jl\}l}\bigl(t,\frac{2\pi}{L}\mathbf{e}_j\bigr)-C^\pm_{k\{kjl\}l}\bigl(t,\frac{2\pi}{L}\mathbf{e}_j\bigr)\bigr)}{\hat{p}_+\hat{t}_-C_{4ll}\bigl(t,\frac{2\pi}{L}\mathbf{e}_j\bigr)}
 \\
 +~&\frac{\hat{p}_+\hat{t}_-\bigl(C^\pm_{j\{j4l\}l}\bigl(t,\frac{2\pi}{L}\mathbf{e}_j\bigr)-C^\pm_{k\{k4l\}l}\bigl(t,\frac{2\pi}{L}\mathbf{e}_j\bigr)\bigr)}{\hat{p}_-\hat{t}_+C_{jll}\bigl(t,\frac{2\pi}{L}\mathbf{e}_j\bigr)}
 \biggr)\end{aligned}&= -\frac{2\pi}{L}E\frac23iR^\goodperp_\pm\,.
\end{align}%
\section{Details of the lattice simulations\label{sec:details}}
For this work we used gauge configurations which have been generated 
using the Wilson gauge action with $N_f=2$ flavors of non-perturbatively 
order $a$ improved Wilson (clover) fermions. A list of the ensembles 
used is shown in table~\ref{table:ListOfLattices}. We used lattices with 
three different inverse couplings $\beta=5.20$, $5.29$, $5.40$, which 
correspond to lattice spacings between $\unit{0.06}{\femto\meter}$ 
and $\unit{0.081}{\femto\meter}$. The pion masses vary between 
$\unit{150}{\mega\electronvolt}$ and
$\unit{500}{\mega\electronvolt}$, with spatial volumes between
$(\unit{1.9}{\femto\meter})^3$ and $(\unit{4.5}{\femto\meter})^3$.

\begin{table}[t]%
\centering%
\begin{tabular}{lccccc}%
 \toprule
 \multicolumn{1}{c}{$\kappa$} & $m_\pi[\mega\electronvolt]$ & 
$m_\pi ^\infty[\mega\electronvolt]$ & Size & $m_\pi L$ & 
$N_{\text{conf}}(\times N_{\text{src}})$\\
 \midrule
 \multicolumn{6}{c}{$\beta = 5.20$, $a = \unit{0.081}{\femto\meter}$, $a^{-1} = \unit{2400}{\mega\electronvolt}$}\\
 \midrule
 $0.13596^\dagger$ & $280$ & 278 & $32^3 \times 64$ & $3.7$ & $1999 (\times 4)$\\
 \midrule
 \multicolumn{6}{c}{$\beta = 5.29$, $a = \unit{0.071}{\femto\meter}$, $a^{-1} = \unit{2800}{\mega\electronvolt}$}\\
 \midrule
 $0.13620^\dagger$ & $422$ & $422$ & $32^3 \times 64$ & $4.8$ & $1998 (\times 2)$\\
 $0.13632$         & $295$ & $290$ & $32^3 \times 64$ & $3.4$ & $1999 (\times 1)$\\
 $0.13632$         & $289$ & $290$ & $40^3 \times 64$ & $4.2$ & $2028 (\times 2)$\\
 $0.13632^\dagger$ & $290$ & $290$ & $64^3 \times 64$ & $6.7$ & $1237 (\times 2)$\\
 $0.13640^\dagger$ & $150$ & $150$ & $64^3 \times 64$ & $3.5$ & $1599 (\times 3)$\\
 \midrule
 \multicolumn{6}{c}{$\beta = 5.40$, $a = \unit{0.060}{\femto\meter}$, $a^{-1} = \unit{3300}{\mega\electronvolt}$}\\
 \midrule
 $0.13640$         & $490$ & $488$ & $32^3 \times 64$ & $4.8$ & $\hphantom{2}982  (\times 2)$\\
 $0.13647^\dagger$ & $426$ & $424$ & $32^3 \times 64$ & $4.2$ & $1999 (\times 2)$\\
 $0.13660$         & $260$ & $259$ & $48^3 \times 64$ & $3.8$ & $2178 (\times 2)$\\
 \bottomrule
\end{tabular}%
\caption{\label{table:ListOfLattices}Ensembles used for this work. 
For each ensemble we give the inverse coupling $\beta$, the hopping parameter
$\kappa$, the pion mass $m_\pi$, the finite volume corrected pion mass 
$m_\pi^\infty$ determined in ref.~\cite{Bali:2016lvx}, the lattice size,
the value of $m_\pi L$, where $L$ is the spatial lattice extent, the
number of configurations $N_{\text{conf}}$ and the number of sources
$N_{\text{src}}$ used on each configuration.
Note that the pion masses have been slightly updated
compared to the numbers in ref.~\cite{Braun:2015axa}. The ensembles 
marked with $^\dagger$ were generated on the QPACE systems of the 
SFB/TRR~55, while the others were generated earlier within the QCDSF 
collaboration.}
\end{table}%
In order to increase the overall statistics
we performed multiple measurements per configuration.
The source positions of these 
measurements were selected randomly to reduce the autocorrelations.
To obtain a better overlap with the ground state we applied Wuppertal 
smearing~\cite{Gusken:1989qx} in the interpolating current $\mathscr{V}_\nu$
using APE smeared gauge links~\cite{Falcioni:1984ei}. 

\begin{figure}[t]%
\centering%
\includegraphics[width=0.5\textwidth]{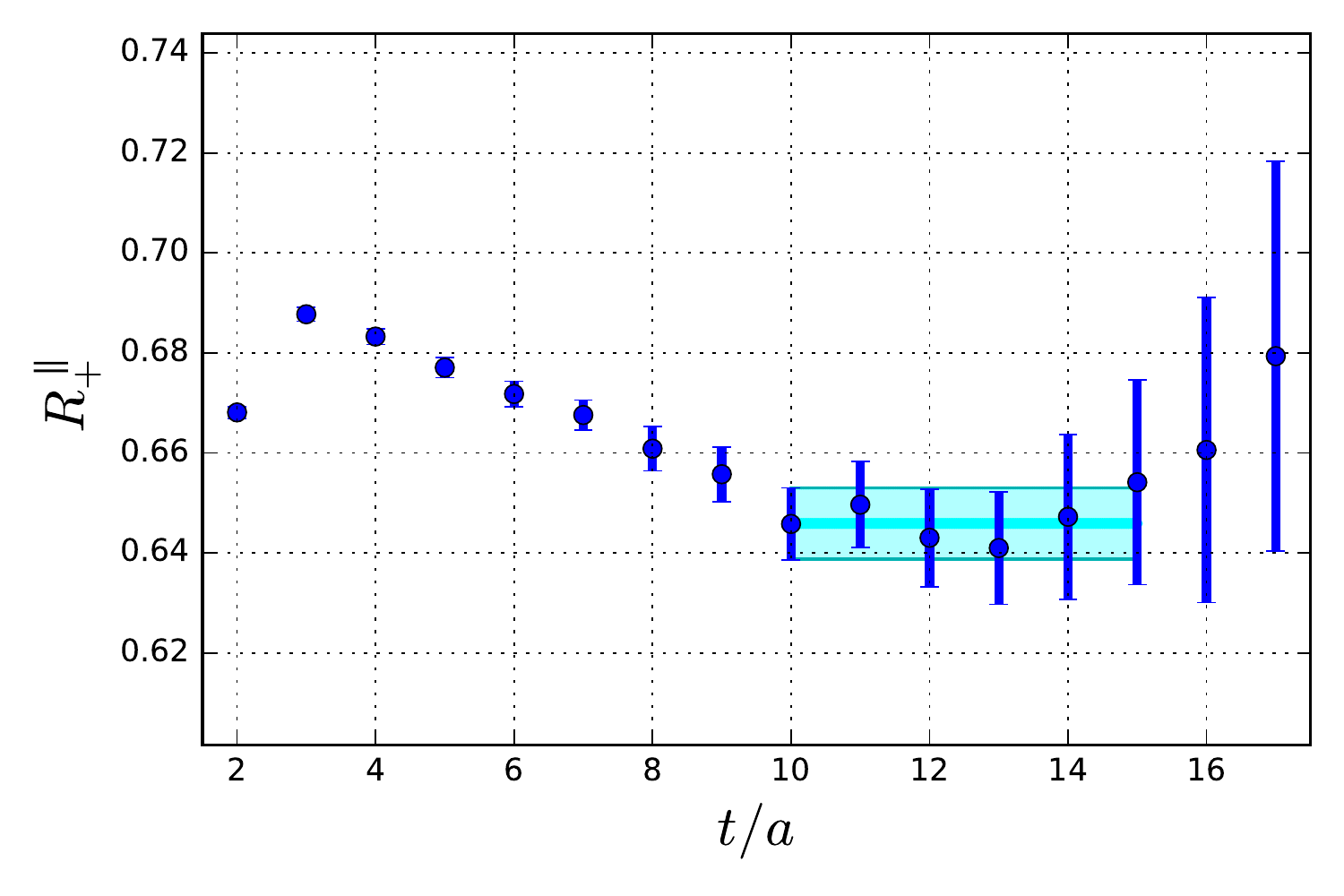}\includegraphics[width=0.5\textwidth]{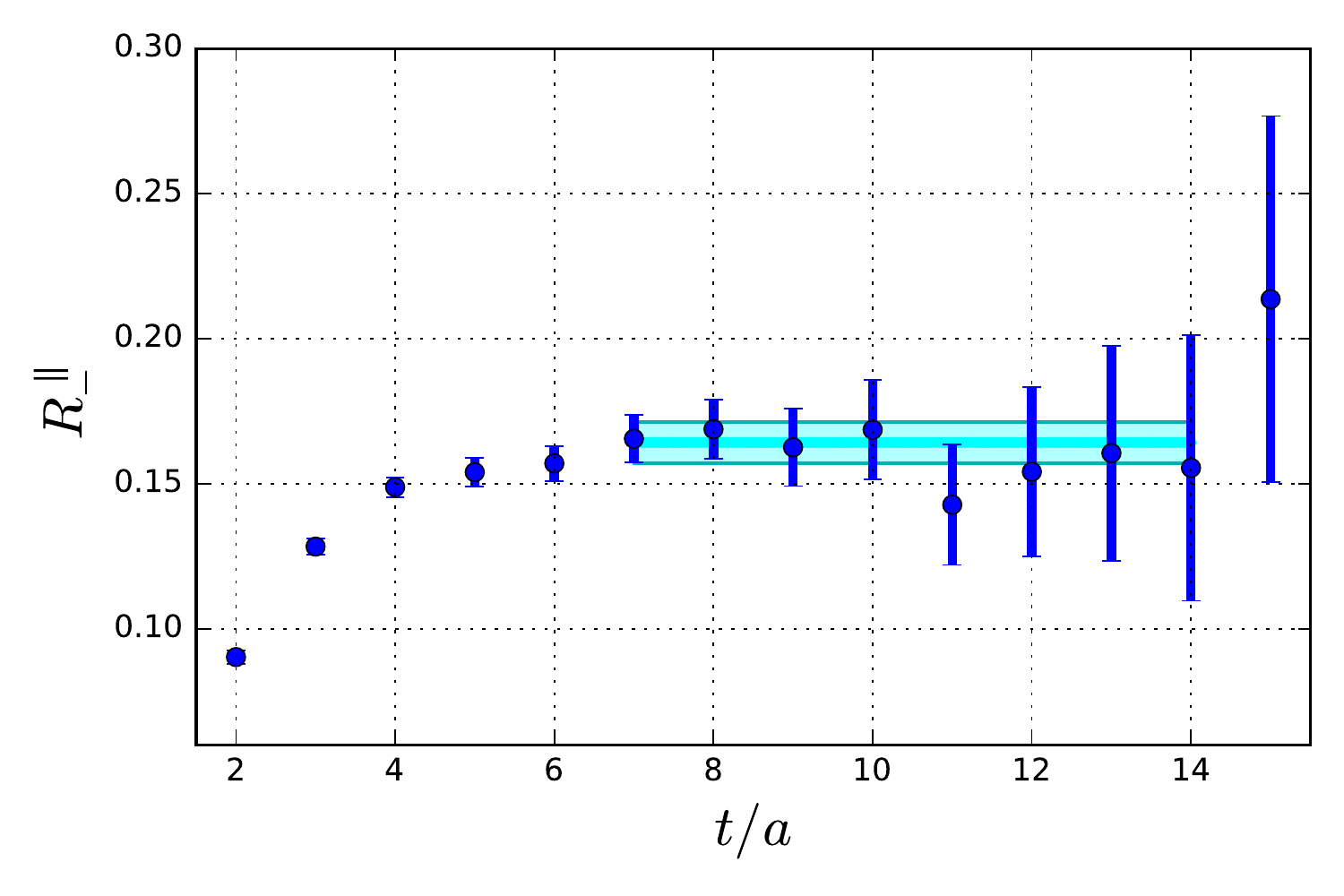}\\%
\caption{\label{fig_range}The data points in these plots show 
$R^{\protect\goodparallel}_\pm$ calculated from the time- 
and momentum-averaged correlation functions according to 
eq.~\eqref{eq-long} on the $\beta=5.29$, $\kappa=0.13632$, $L=32a$, $T=64a$ 
ensemble. The cyan-colored bar indicates the fitted value of 
$R^{\protect\goodparallel}_\pm$, the error and the fitting range.}
\end{figure}

For the statistical analysis we generate $1000$ bootstrap samples per 
ensemble using a binsize of $4$ to further eliminate autocorrelations. 
For the purpose of maximizing the statistics of the second moments, 
we average for each bootstrap sample over all suitable combinations of 
2-point functions, all possible momentum directions
as well as over forward and backward running states as pointed 
out in eqs.~\eqref{eq-long} and~\eqref{eq-trans}.
In order to reduce contributions from excited states the choice of the 
starting point of the fit range is important.
As an example, figure~\ref{fig_range} demonstrates that, with increasing 
source-sink distance, the excited states fall below the noise
and  plateaus of the correlation functions for $R_\pm ^\goodparallel$ 
emerge. The starting time $t_{\text{start}}$ is then chosen in such a 
way that fits with even larger starting times no longer show any 
systematic trend in the fitted values.
Multi-state fits (over larger fit ranges) yield consistent results. 
\section{Renormalization\label{sec:reno}}
Having computed the bare values of the second DA moments, we are left 
with the task of renormalizing these bare quantities to obtain 
results in the standard continuum $\MSbar$ scheme, which is used in 
the perturbative calculations of the exclusive reactions discussed 
in the introduction. As already 
mentioned above, our bare operators are chosen such that there is only 
mixing between the respective $+$ and $-$ operator multiplets, so we 
have to determine $2 \times 2$ mixing matrices such that 
\begin{subequations}
\begin{align}
\mathcal O_{\MSbar}^- &= Z_{11} \mathcal O^- + Z_{12} \mathcal O^+ \,, \\
\mathcal O_{\MSbar}^+ &= Z_{22} \mathcal O^+ \,.
\end{align}
\end{subequations}
One then obtains for the second moments of the DAs in the $\MSbar$ scheme
\begin{align}
a_{2,\MSbar}^{\goodparallel,\goodperp} &= \tfrac{7}{12} 
\bigl[ 5\zeta^{\goodparallel,\goodperp}_{11} R^{\goodparallel,\goodperp}_- 
+ \bigl( 5 \zeta^{\goodparallel,\goodperp}_{12} 
- \zeta^{\goodparallel,\goodperp}_{22} \bigr) 
R^{\goodparallel,\goodperp}_+  \bigr]\,,\\
\langle \xi^2 \rangle^{\goodparallel,\goodperp}_{\MSbar} &= 
\zeta^{\goodparallel,\goodperp}_{11} R^{\goodparallel,\goodperp}_- 
+ \zeta^{\goodparallel,\goodperp}_{12} R^{\goodparallel,\goodperp}_+\,,\\
 \langle 1^2 \rangle^{\goodparallel,\goodperp}_{\MSbar} &= 
\zeta^{\goodparallel,\goodperp}_{22} R^{\goodparallel,\goodperp}_+\,,
\end{align}
where
\begin{equation}
 \zeta^{\goodparallel}_{ij} = \frac{Z^{\goodparallel}_{ij}}{Z_V} \quad , \quad
 \zeta^{\goodperp}_{ij} = \frac{Z^{\goodperp}_{ij}}{Z_T}
\end{equation}
with the renormalization factors $Z_V$ and $Z_T$ of the vector and the tensor
currents, respectively.
Note that one cannot expect $\zeta^{\goodparallel,\goodperp}_{22}$ 
to be equal to one, since the Leibniz rule holds on the lattice 
only up to discretization artefacts.

We want to evaluate the renormalization and mixing coefficients 
non-perturbatively on the lattice employing a variant of the RI${}^\prime$-MOM 
scheme, because lattice perturbation theory is not sufficiently reliable.
Since forward matrix elements of the $+$ operators vanish in the 
continuum limit, we cannot work with the momentum geometry of the 
original RI${}^\prime$-MOM scheme but must use a kind of RI${}^\prime$-SMOM 
scheme~\cite{Sturm:2009kb}. We follow exactly the same renormalization 
procedure as in our investigation of the pion DA~\cite{Braun:2015axa}.
Thus, we need the $\MSbar$ vertex functions of our operators
in order to convert the results from our SMOM scheme to the $\MSbar$ scheme.
While these are known to two loops in the longitudinal case, i.e.,
for the operators \eqref{def:bareops_long}, see ref.~\cite{Gracey:2011zg}, 
as well as for the currents \eqref{def:currents}, see ref.~\cite{Gracey:2011fb},
the corresponding results for operators with derivatives involving the matrix 
$\sigma_{\mu \nu}$, e.g., the operators \eqref{def:bareops_trans},
are not yet available in the literature. Therefore we discuss the latter
case, the so-called transversity operators, in appendix~\ref{app_trans}. 

In the end, we determine the matrix $Z(a,\mu_0)$ (and analogously
$\zeta(a,\mu_0)$) at the reference scale
$\mu_0 = \unit{2}{\giga\electronvolt}$ by fitting the chirally 
extrapolated Monte Carlo results
$Z(a,\mu)_{\mathrm {MC}}$ with the expression
\begin{equation} \label{eq:fitfun}
Z(a,\mu)_{\mathrm {MC}} = W(\mu, \mu_0) Z(a,\mu_0) + A_1 a^2 \mu^2 
  + A_2 (a^2 \mu^2)^2 + A_3 (a^2 \mu^2)^3 \,,
\end{equation}
where the three matrices $A_i$ parametrize the lattice artefacts and
$W(\mu, \mu_0)$ describes the running of $Z$ in the three loop 
approximation of continuum perturbation theory. 

Ignoring the very small statistical errors, we estimate the 
much more important systematic uncertainties of $Z(a,\mu_0)$
by performing a number of 
fits, where exactly one element of the analysis is varied at a time.
More precisely, we choose as representative examples for fit intervals 
$\unit{4}{\giga\electronvolt}^2 < \mu^2 < \unit{100}{\giga\electronvolt}^2$
and 
$\unit{2}{\giga\electronvolt}^2 < \mu^2 < \unit{30}{\giga\electronvolt}^2$,
and we use the expressions for the conversion functions 
with $n_{\mathrm {loops}}= 1,2$. For the parametrization of the lattice 
artefacts we either take the complete expression in eq.~(\ref{eq:fitfun})
or we set $A_3=0$. Finally, we consider values for $r_0$ and 
$r_0 \Lambda_{\MSbar}$ corresponding to the results given in 
ref.~\cite{Fritzsch:2012wq}. The various fit possibilities are compiled
in table~\ref{table:fit_choices}.

\begin{table}[t] 
\renewcommand{\arraystretch}{1.2}
\centering
\begin{tabular}{cccccc}
 \toprule
Fit    & Fit interval & $n_{\mathrm {loops}}$ & Lattice   &  $r_0$   & $r_0 \Lambda_{\MSbar}$ \\
number & (in $\giga\electronvolt\squared$)  &                 & artefacts & (in $\femto\meter$)  &                  \\ \midrule
 1 & $4 \, < \mu^2 < 100 \,$ & 2 & 
 $A_3 \neq 0$ & $0.50 \, $ & 0.789 \\
 2 & $2 \,  < \mu^2 < 30\hphantom{0} \, $ & 2 & 
 $A_3 \neq 0$ & $0.50 \, $ & 0.789 \\
 3 & $4 \, < \mu^2 < 100 \, $ & 1 & 
 $A_3 \neq 0$ & $0.50 \,$ & 0.789 \\
 4 & $4 \, < \mu^2 < 100 \, $ & 2 & 
 $A_3=0$ & $0.50 \,$ & 0.789 \\
 5 & $4 \, < \mu^2 < 100 \,$ & 2 & 
 $A_3 \neq 0$ & $0.49 \,$ & 0.789 \\
 6 & $4 \, < \mu^2 < 100 \, $ & 2 & 
 $A_3 \neq 0$ & $0.50 \,$ & 0.737 \\
 \bottomrule
\end{tabular}
\caption{\label{table:fit_choices}Choices for the fits of the 
renormalization and mixing coefficients.} 
\renewcommand{\arraystretch}{1.0}
\end{table}

As in the case of the pion DA, the largest effect comes from the 
variation of $n_{\mathrm {loops}}$. In order to obtain our
final numbers for the second moments of the DAs we extract them from 
the bare data $R^{\goodparallel,\goodperp}_\pm$ using each of these
sets of values for $\zeta_{11}$, $\zeta_{12}$ and $\zeta_{22}$.    
So we have six results for each of our gauge field ensembles.
As our central values we take the results from Fit 1. 
Defining $\delta_i$ as the difference between
the result obtained with the $\zeta$s from Fit $i$ and the result 
determined with the $\zeta$s from Fit $1$, we estimate the  
systematic uncertainties due to the renormalization factors 
as $\sqrt{\delta_2^2 + (0.5 \cdot \delta_3)^2 + \delta_4^2 + 
          \delta_5^2 + \delta_6^2}$. Here we have multiplied $\delta_3$
by $1/2$, because going from two loops to three or more loops in the 
perturbative conversion functions is expected to lead to a smaller change than
going from one loop to two loops. This should amount to a rather 
conservative error estimate.
The renormalization factors $Z_V$ and $Z_T$ needed for the evaluation of
$f_\rho^{\vphantom{T}}$ and $f_\rho^T$, respectively, are calculated in the same way.

\section{Data analysis}%

\begin{figure}[t]%
\centering%
\includegraphics[width=0.5\textwidth]{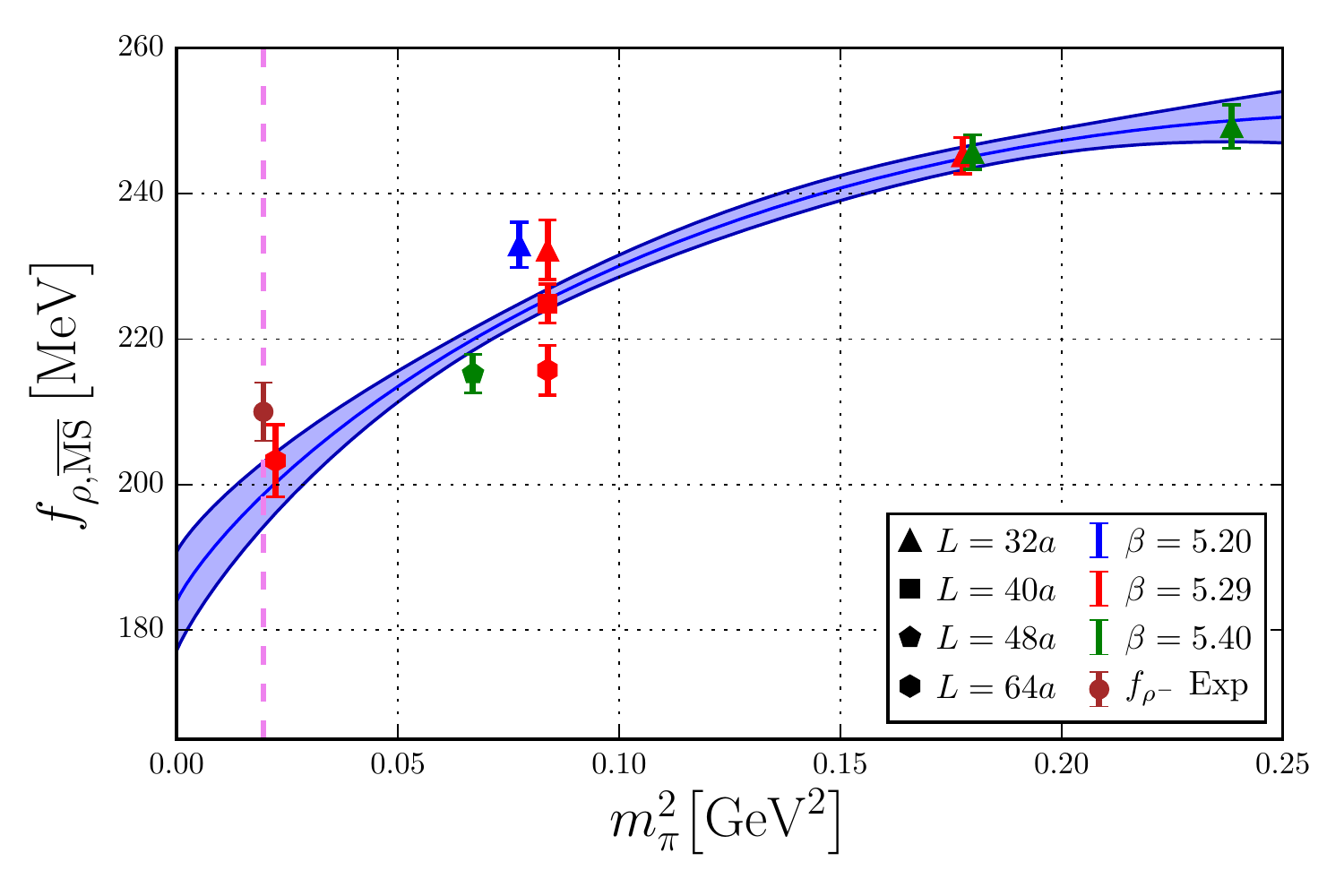}\includegraphics[width=0.5\textwidth]{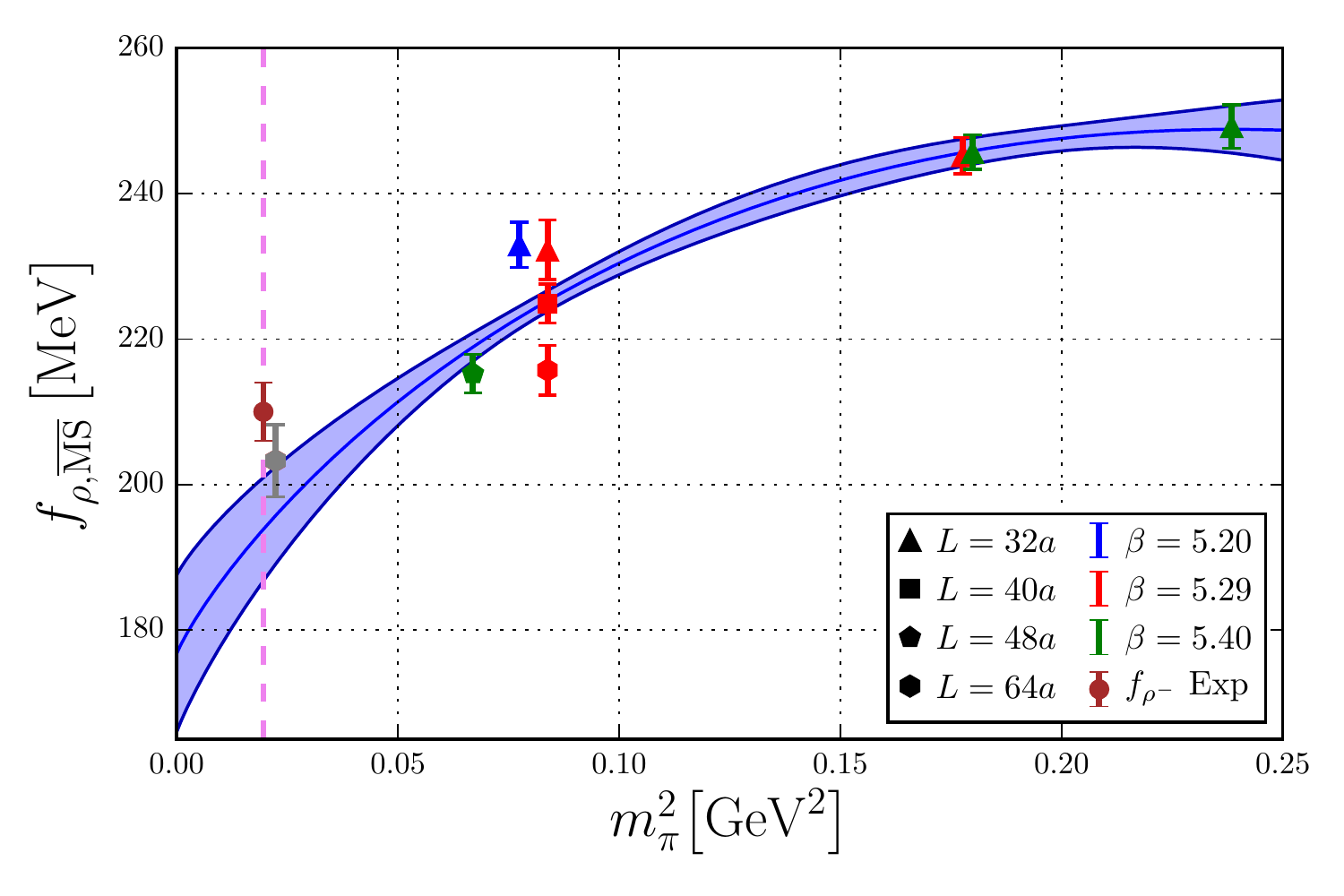}\\%
\includegraphics[width=0.5\textwidth]{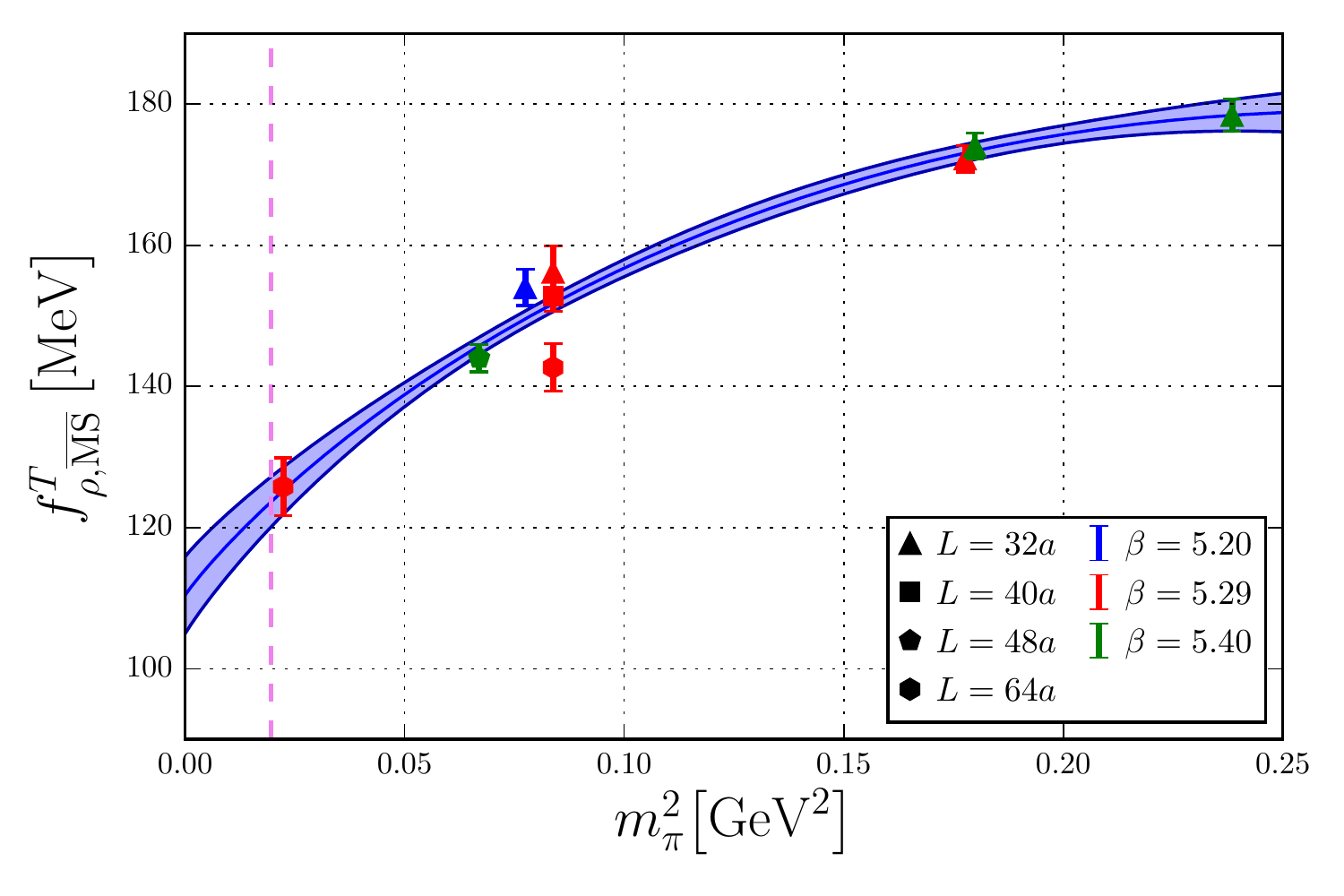}\includegraphics[width=0.5\textwidth]{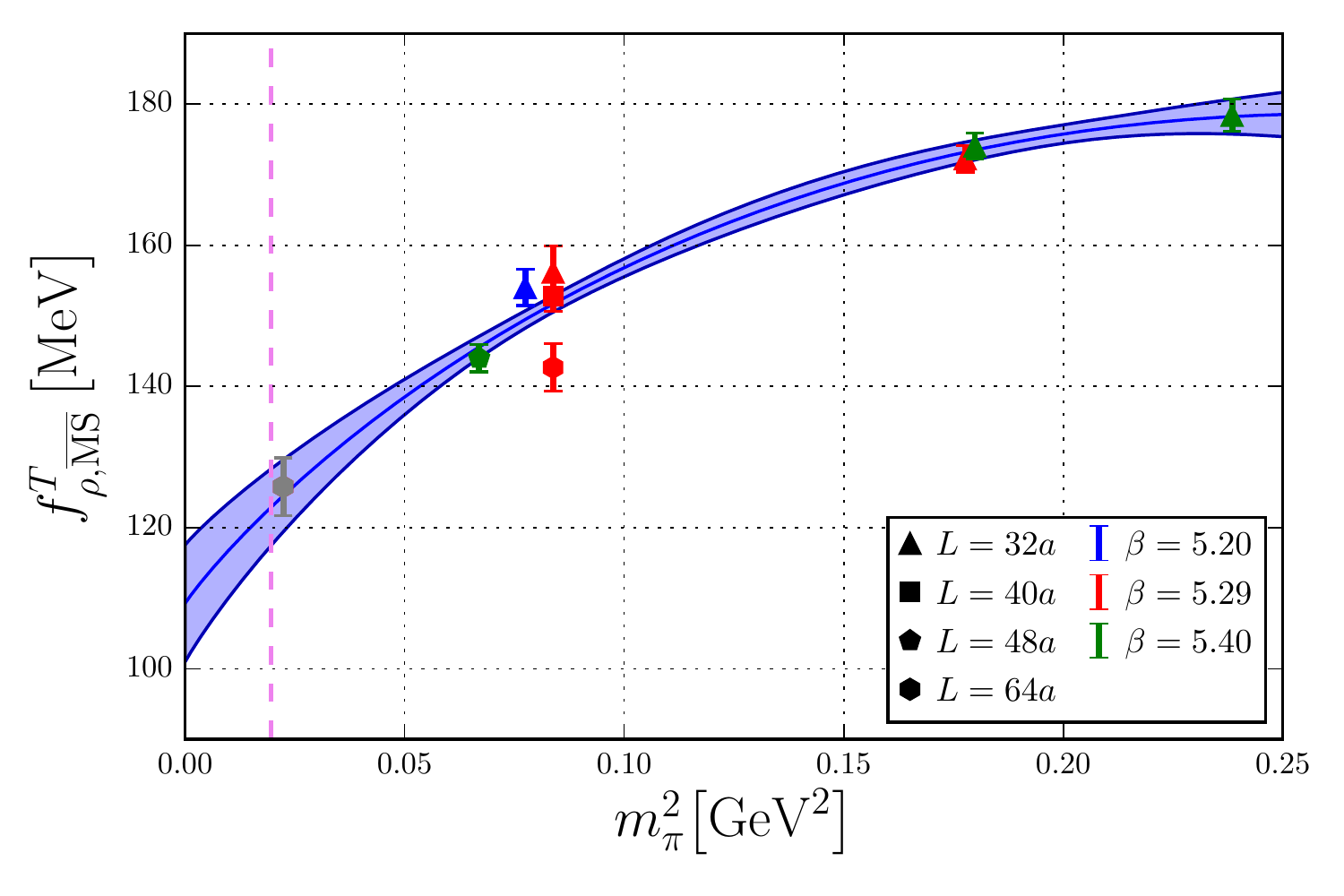}\\%
\includegraphics[width=0.5\textwidth]{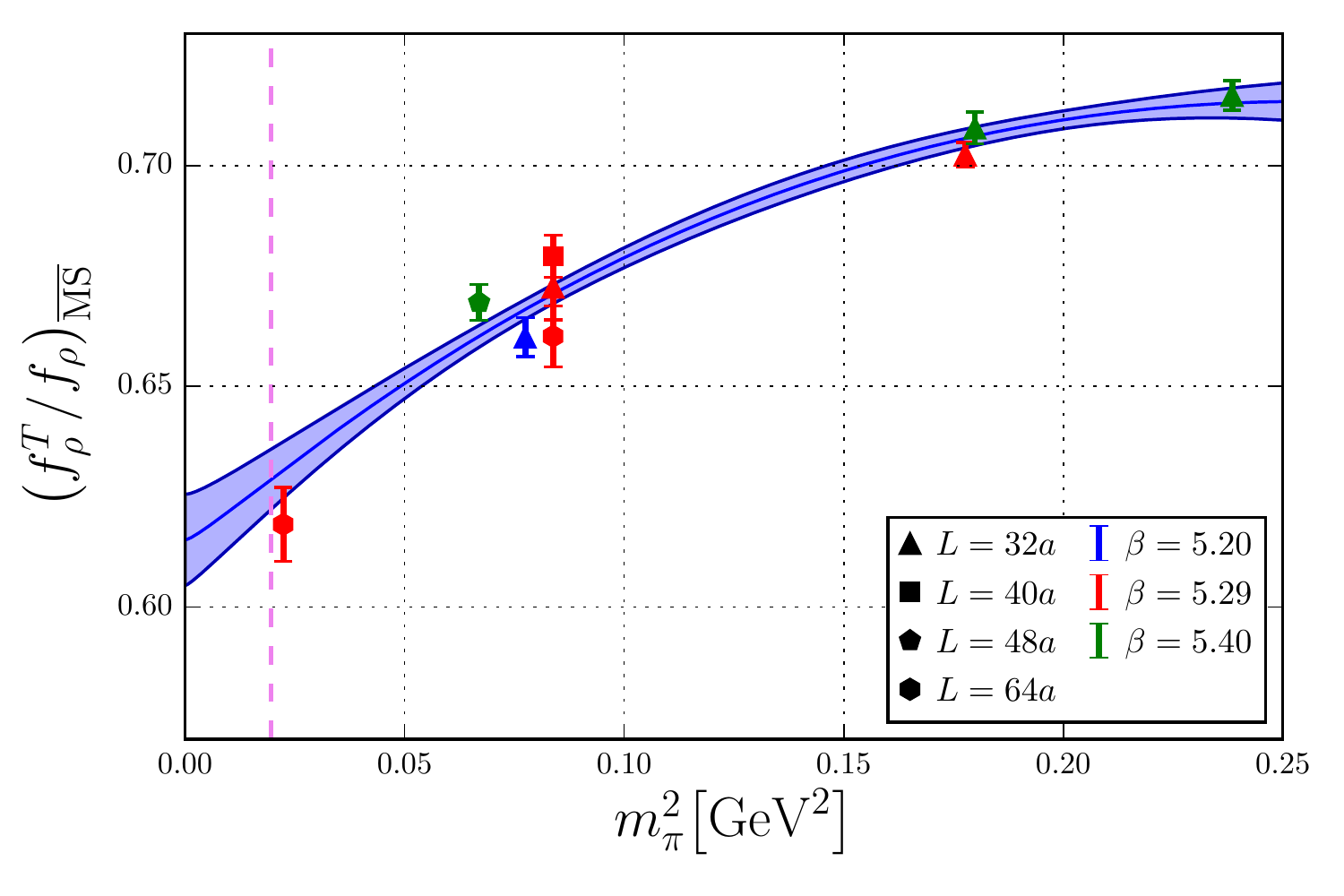}\includegraphics[width=0.5\textwidth]{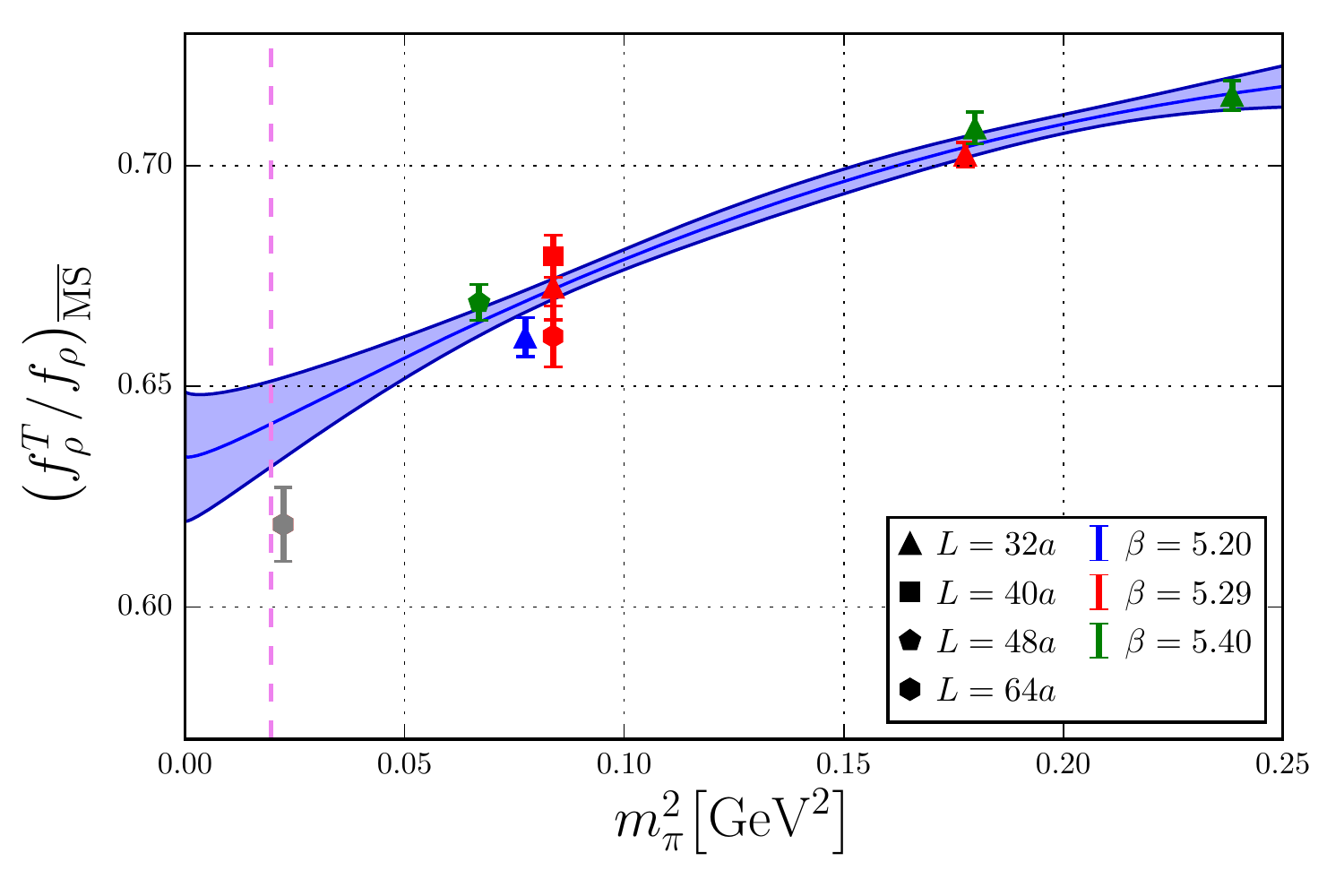}%
\caption{\label{figure_chiral_extrapolation_norm}ChPT fits using eqs.~\eqref{eq_chpt_fits} for the decay constants $f_\rho^{\protect\vphantom{T}}$, $f_\rho^T$ and their ratio, including (left) and excluding (right) the data point at $m_\pi=\unit{150}{\mega\electronvolt}$. The violet dashed line indicates the position of the physical pion mass. The band indicates the one sigma statistical error.}%
\end{figure}%
\begin{figure}[t]%
\centering%
\includegraphics[width=0.5\textwidth]{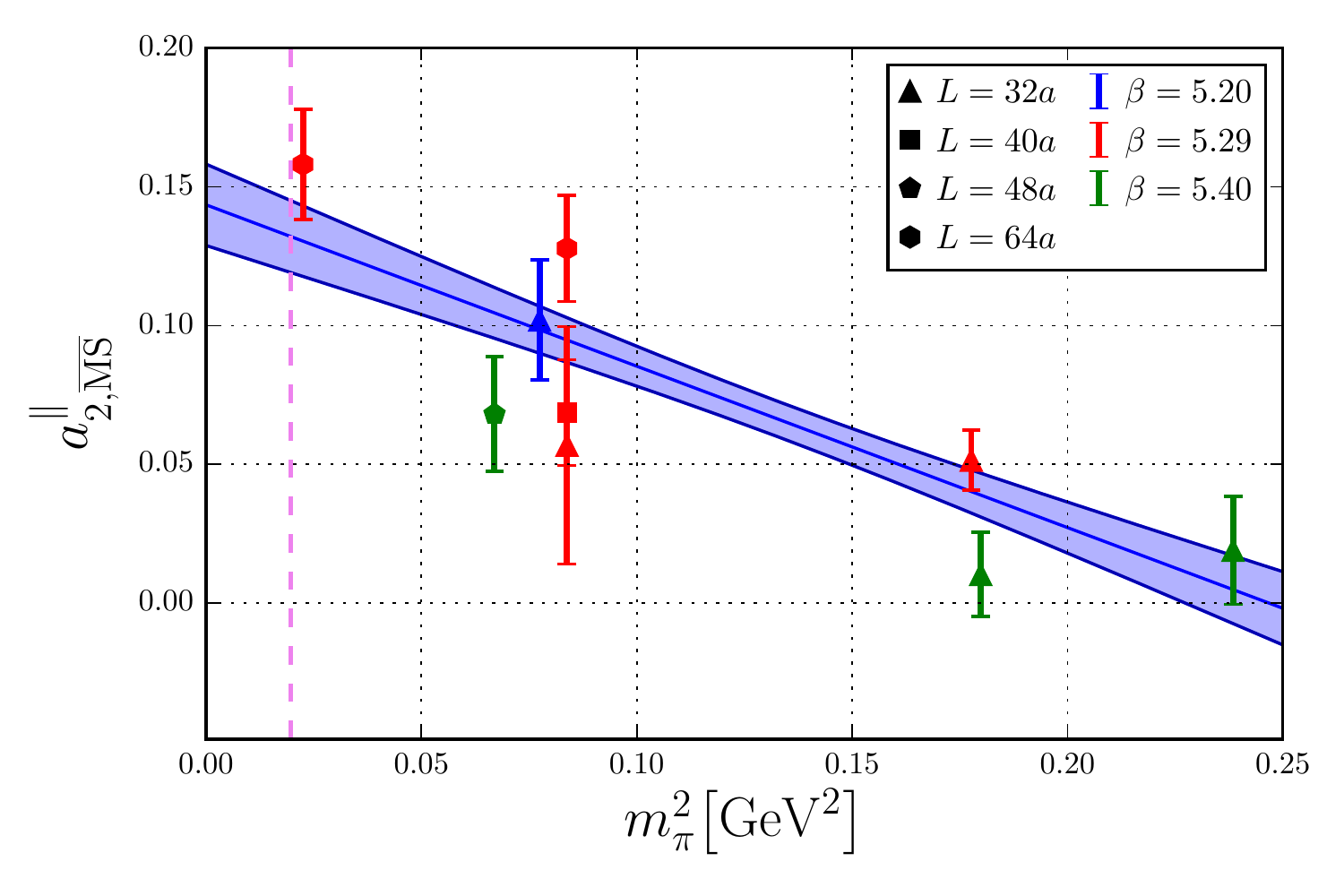}\includegraphics[width=0.5\textwidth]{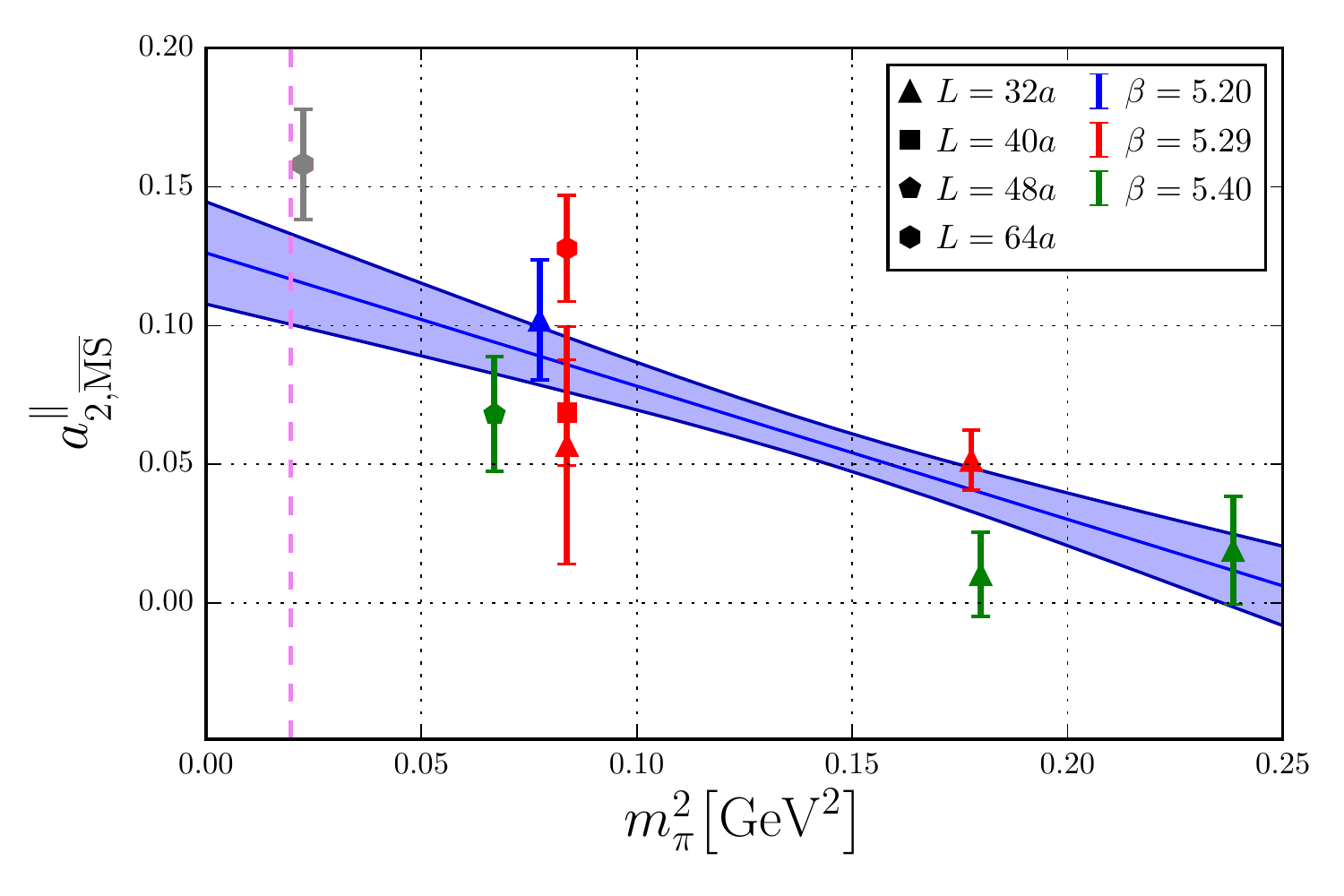}\\%
\includegraphics[width=0.5\textwidth]{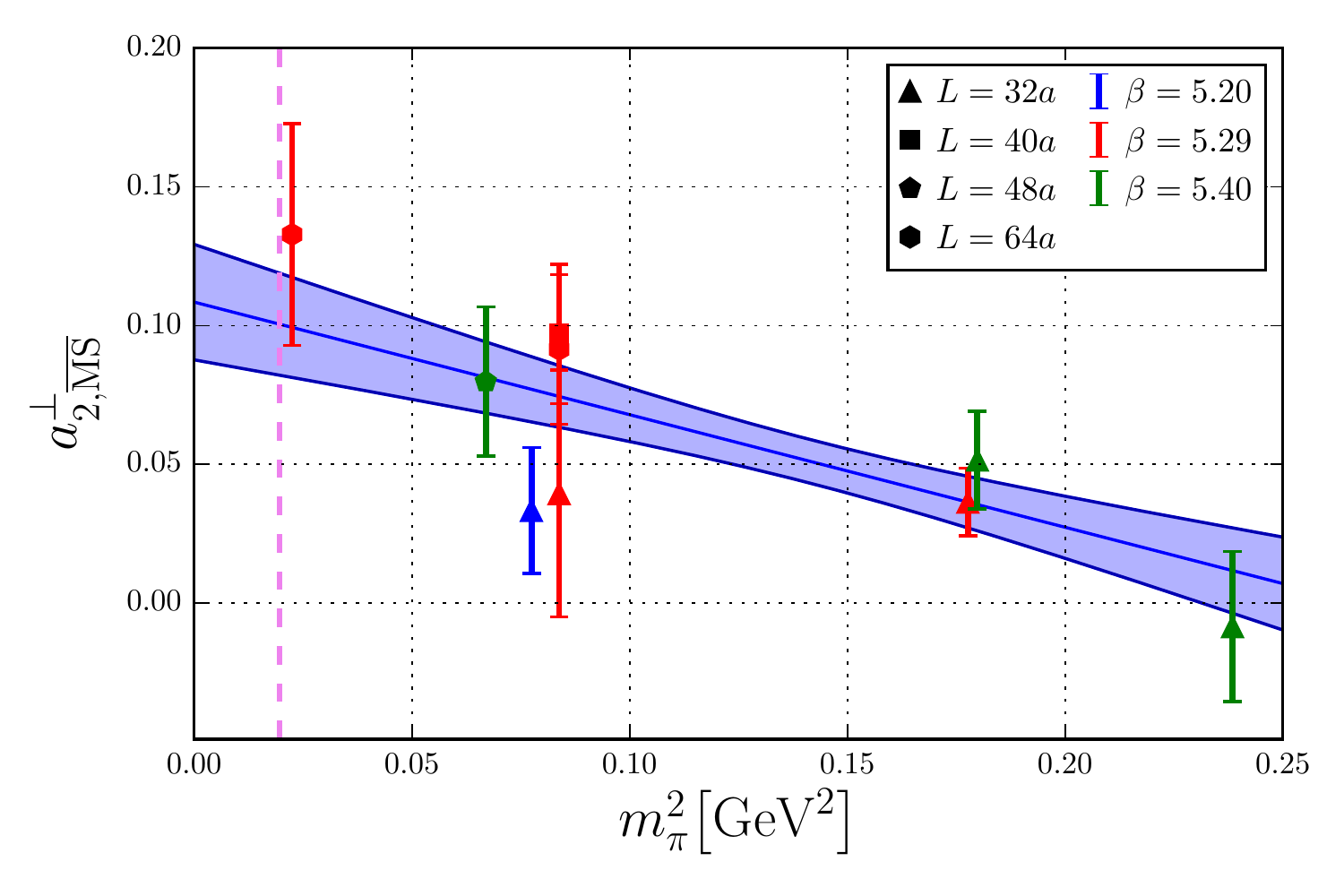}\includegraphics[width=0.5\textwidth]{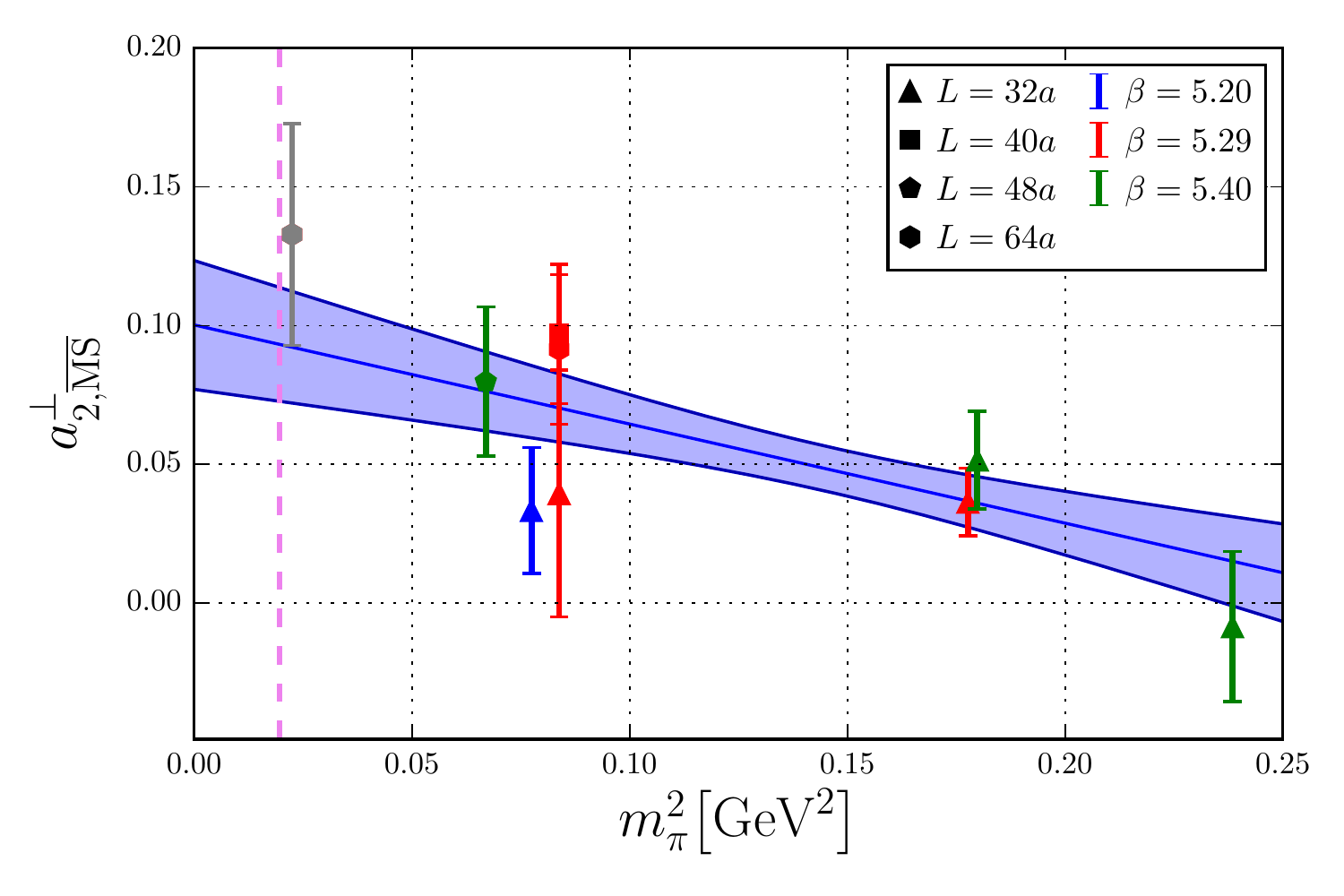}%
\caption{\label{figure_chiral_extrapolation_mom}Linear fits for the 
second Gegenbauer moments $a_{2,\MSbar}^{\protect\goodparallel}$, 
$a_{2,\MSbar}^{\protect\goodperp}$ of the linearly and the transversely 
polarized leading twist distribution amplitudes, including (left) and 
excluding (right) the data point at $m_\pi=\unit{150}{\mega\electronvolt}$. 
The violet dashed line indicates the position of the physical pion mass. 
The band indicates the one sigma statistical error.}%
\end{figure}%

From the bare values of $f_\rho$ etc.\ we obtain renormalized results 
in the $\MSbar$ scheme with the
help of our renormalization (and mixing) coefficients on each of our 
gauge field ensembles. 
With the range of ensembles available (see table~\ref{table:ListOfLattices}) 
we are able to study the pion mass dependence and, to only a limited 
extent, volume and discretization effects. 
Since our lattice spacings do not vary that much, a continuum extrapolation 
cannot be attempted. Moreover, the impact of the 
finite lattice size on matrix elements of possibly unstable states 
is not straightforward. So we take into account results from all lattice 
spacings and volumes for our final numbers. 

Considering the pion mass dependence, we make use of Chiral Perturbation Theory (ChPT)
for vector mesons~\cite{Ecker:1988te,Ecker:1989yg,Jenkins:1995vb} to 
obtain the one loop extrapolation formulae for the decay constants

\begin{subequations}\label{eq_chpt_fits}%
\begin{align}%
 \operatorname{Re} f_\rho &= f_\rho^{(0)}\biggl(1-\frac{m_\pi^2}{16\pi^2F_\pi^2}\log\biggl(\frac{m_\pi^2}{\mu_\chi^2}\biggr)\biggr) + f_\rho^{(2)}m_\pi^2 + f_\rho^{(3)}m_\pi^3 + \mathcal O(m_\pi^4)\,,\label{eq_chpt_fits_long}\\
 \operatorname{Re} f_\rho^T &= f_\rho^{T(0)}\biggl(1-\frac{m_\pi^2}{32\pi^2F_\pi^2}\log\biggl(\frac{m_\pi^2}{\mu_\chi^2}\biggr)\biggr) + f_\rho^{T(2)}m_\pi^2 + f_\rho^{T(3)}m_\pi^3 + \mathcal O(m_\pi^4)\,,\label{eq_chpt_fits_trans}\\
 \operatorname{Re} \frac{f_\rho^T}{f_\rho} &= \delta f_\rho^{(0)}\biggl(1+\frac{m_\pi^2}{32\pi^2F_\pi^2}\log\biggl(\frac{m_\pi^2}{\mu_\chi^2}\biggr)\biggr) + \delta f_\rho^{(2)}m_\pi^2 + \delta f_\rho^{(3)}m_\pi^3 + \mathcal O(m_\pi^4) \,.
\label{eq_chpt_fits_ratio}
\end{align}%
\end{subequations}%
Details on the ChPT calculation are given in appendix~\ref{app_chpt}. 
For $2 m_\pi < m_\rho$, i.e., below the decay threshold 
this infinite-volume calculation yields 
complex numbers. However, as we neglect instability effects in our 
lattice computation, which is necessarily done on finite volumes, we 
use only the real part to fit the mass dependence of our data.
The pion decay constant $F_\pi$ is kept fixed at its physical value 
$\unit{92.4}{\mega\electronvolt}$, and the chiral renormalization scale
$\mu_\chi$ is chosen to be $\unit{775}{\mega\electronvolt}$.

Estimates within ChPT suggest that the third-order term $\propto m_\pi^3$
is not negligible for most of our masses (see appendix~\ref{app_chpt}). 
Our data confirm this expectation --- the third order term is
required in order to fit over the full range of pion
masses. Consistent fits are obtained including only second order terms
for $m_\pi <\unit{300}{\mega\electronvolt}$, however, we have, essentially, 
only two pion masses in this range. Alternatively, one can ignore the 
information from ChPT and perform polynomial fits, i.e., drop the 
logarithmic term in the fit functions \eqref{eq_chpt_fits}. This yields 
very similar results. We expect that a fit including the larger pion 
masses will yield more reliable numbers than simply taking the values at 
$m_\pi = \unit{150}{\mega\electronvolt}$ as our final results because, 
in particular, the lattice used at this pion mass is relatively small. 

In order to get at least some idea of the influence of the
instability of the $\rho$ we perform two kinds of fits, including all masses
or excluding the results at $m_\pi=\unit{150}{\mega\electronvolt}$,
which should suffer most from the decay.   
The resulting ChPT extrapolations for the normalization constants are shown 
in figure~\ref{figure_chiral_extrapolation_norm}.
Note that the extrapolated values at the physical point 
are reasonably consistent with the data at the lowest pion mass.

For the second moments of vector meson distribution amplitudes
(see figure~\ref{figure_chiral_extrapolation_mom}) no ChPT calculations 
are available. It is known that these quantities for pseudoscalar 
mesons~\cite{Chen:2003fp,Chen:2005js} and the nucleon~\cite{Wein:2015oqa}
do not contain chiral logarithms in leading one-loop order. The 
reasons are rather generic and may apply 
to vector mesons as well. Therefore we stick to simple linear 
fits in $m_\pi^2$ depicted in figure~\ref{figure_chiral_extrapolation_mom}.
There is no discernible dependence on the lattice spacing.
Errors stemming from the renormalization constants are not included 
in figures~\ref{figure_chiral_extrapolation_norm}--\ref{figure_chiral_extrapolation_mom}.
We perform an extrapolation for every choice given in 
table~\ref{table:fit_choices} and compute the error of the extrapolated result
at the physical point caused by the uncertainties of the renormalization
factors from the differences of the extrapolated numbers as indicated 
at the end of section~\ref{sec:reno}.

\begin{figure}[t]
\centering
\includegraphics[width=0.5\textwidth]{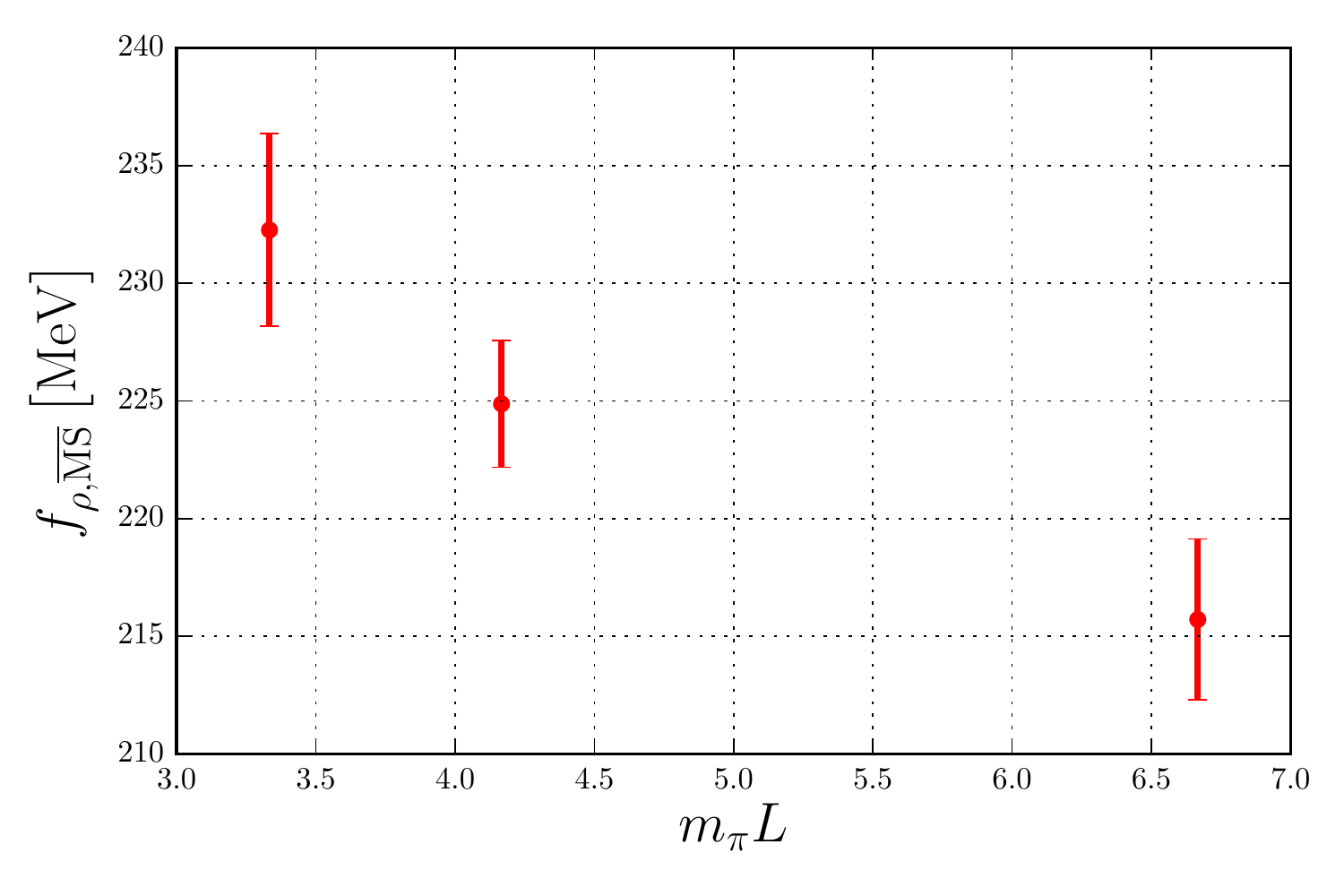}\includegraphics[width=0.5\textwidth]{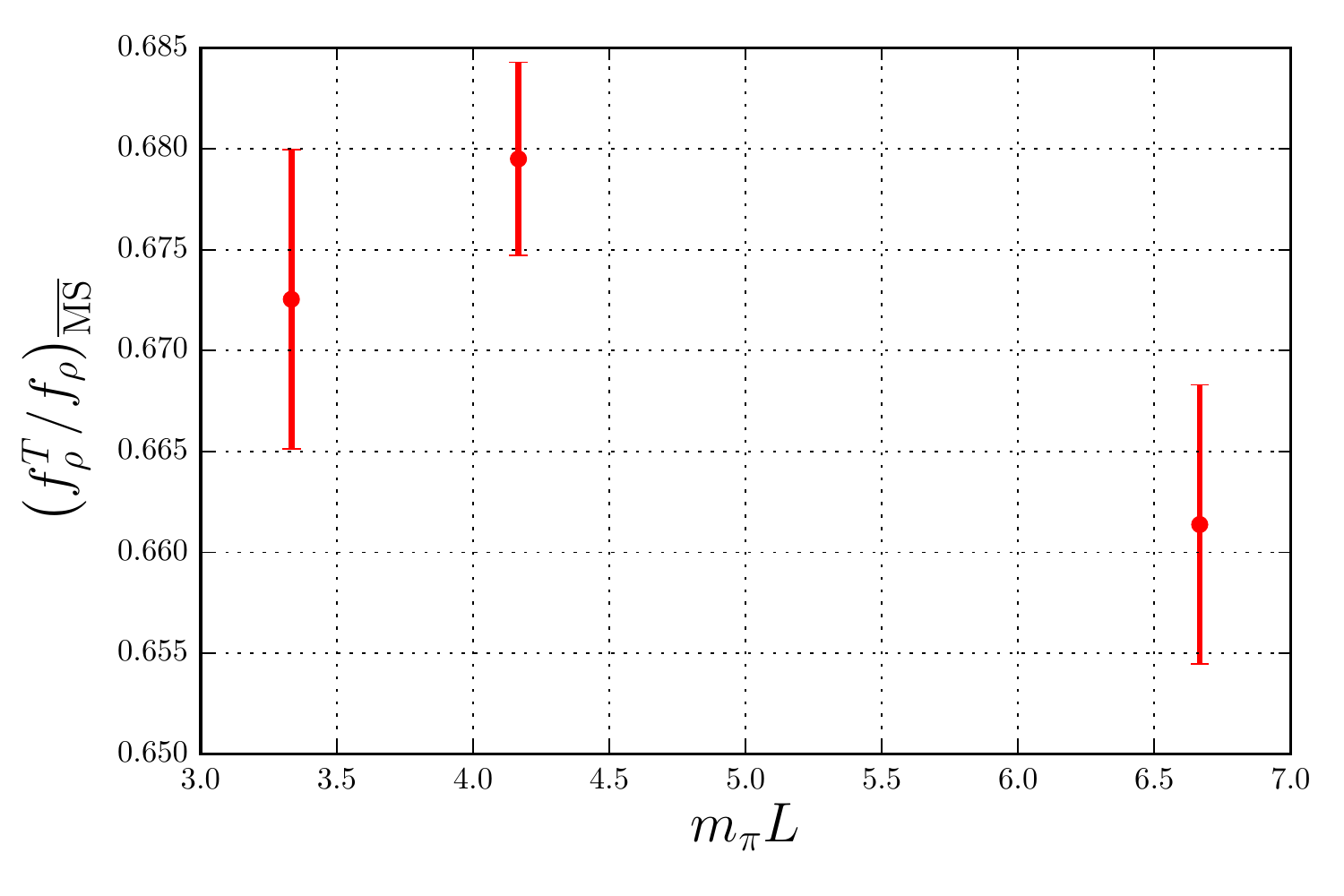}
\caption{\label{figure_volume_extrapolation}Volume dependence of $f_\rho$
(left panel) and $f_\rho^T/f_\rho^{\protect\vphantom{T}}$ (right panel) 
at $\beta=5.29$, $\kappa=0.13632$.}
\vspace{1cm}
\includegraphics[width=0.5\textwidth]{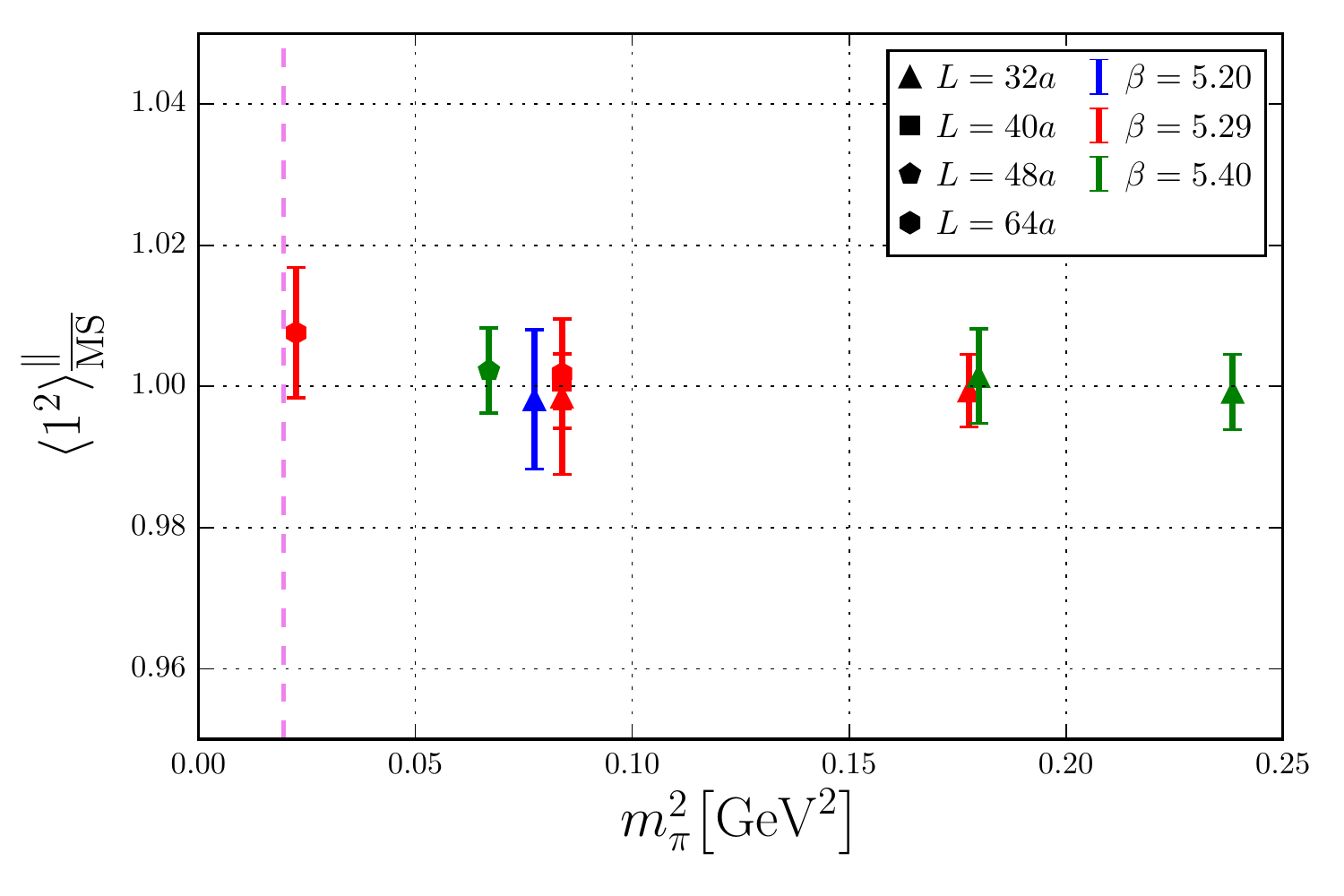}\includegraphics[width=0.5\textwidth]{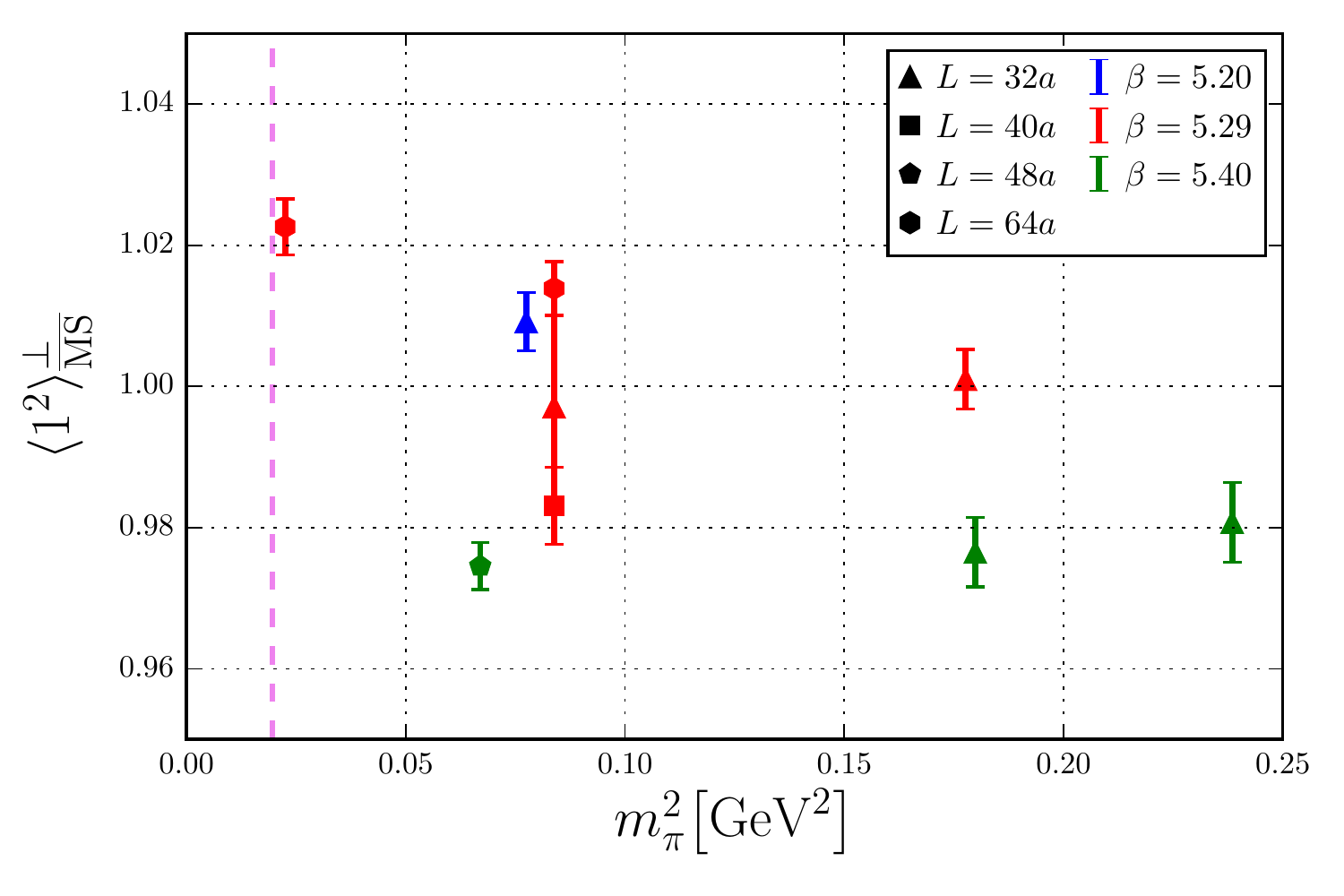}%
\caption{\label{figure_eins}The left and the right plot show 
$\langle 1^2 \rangle^{\protect\goodparallel}_{\MSbar}$ and
$\langle 1^2 \rangle^{\protect\goodperp}_{\MSbar}$, respectively. 
The violet dashed line indicates the position of the physical pion mass.}%
\end{figure}

Although our data do not allow us to study finite-size and discretization 
effects systematically, we can make some observations.
Considering volume effects, for $\beta = 5.29$, $\kappa = 0.13632$ we 
have ensembles with three different volumes at our disposal 
($m_\pi L = 3.4\text{--}6.7$). The effects for the decay constants are sizable, 
see figures~\ref{figure_chiral_extrapolation_norm} 
and~\ref{figure_volume_extrapolation}.
Unlike the well-known cases of pseudoscalar meson and baryon masses, the 
chiral extrapolations 
cannot be converted directly to predictions for the leading large-volume 
behavior. The problematic contributions cancel, however, in the
ratio of the decay constants $f_\rho^T/f_\rho^{\vphantom{T}}$, so that it is straightforward 
to compute the leading finite-volume corrections for this ratio (see  
appendix~\ref{app_chpt}). It turns out that the corrections are numerically 
tiny so that from the ChPT analysis one expects that finite-volume effects 
for  $f_\rho^T/f_\rho^{\vphantom{T}}$ are much smaller than for the couplings themselves. 
This expectation is in agreement with our data, as shown in figure~\ref{figure_volume_extrapolation}: 
The finite-volume effects for the ratio $f_\rho^T/f_\rho^{\vphantom{T}}$ (right panel) are 
considerably smaller than for the vector coupling $f_\rho$ itself (left panel).
Since in phenomenological studies of hard reactions $f_\rho$ will always
be set to the experimental value, the ratio $f_\rho^T/f_\rho^{\vphantom{T}}$, which is not
experimentally accessible, is a much more interesting quantity. So 
we do not perform an infinite-volume extrapolation 
for $f_\rho$ and use this measurement mostly for normalization purposes 
(e.g.\ computing the second moments). On the other hand,
the observed volume dependence of $f_\rho^T/f_\rho^{\vphantom{T}}$ is small compared to 
the statistical errors and an infinite-volume extrapolation would not have 
any significant effect. 

One can see from figure~\ref{figure_chiral_extrapolation_mom} that the 
second moments tend to increase with the spatial volume, however, less 
significantly than for the normalization constants and the data points 
have comparatively much larger error bars. 
As mentioned above, the ChPT analysis of the second moments is not 
available but the corresponding quantities for stable hadrons have no 
leading chiral logarithms and a very mild finite-volume dependence. 
We have checked that excluding the 
smallest-volume lattice with $m_\pi L  = 3.4$ from the fits does not have any 
noticeable influence on our results.  

Discretization errors are notoriously difficult to control.
A certain insight can be obtained looking 
at the quantities 
$\langle 1^2 \rangle^{\protect\goodparallel}_{\MSbar}$ and
$\langle 1^2 \rangle^{\protect\goodperp}_{\MSbar}$, which indicate the
violation of the Leibniz rule at finite lattice spacing. In the continuum
limit they should equal one for all pion masses. Results for all
ensembles are plotted in figure~\ref{figure_eins}. Again only statistical
errors are shown, the uncertainties resulting from the renormalization
coefficients are much smaller.
While $\langle 1^2 \rangle^{\protect\goodparallel}_{\MSbar}$ equals one
within the statistical errors with a maximal deviation of about 1\%, 
we observe deviations from one of up to 2\% for
$\langle 1^2 \rangle^{\protect\goodperp}_{\MSbar}$. Note that these
deviations are noticeably smaller than what we found in the case of the 
pion~\cite{Braun:2015axa}.

\section{Results and conclusion}%

\begin{table}[t]
\centering%
\begin{tabular}{cccccc}%
\toprule
 & $f_\rho [\mega\electronvolt]$ & $f_\rho^T [\mega\electronvolt]$ & 
$f_\rho^T/f_\rho^{\vphantom{T}}$ & $a_2^\goodparallel$ & $a_2^\goodperp$ \\
\midrule
analysis 1 & $199(4)(1)$ & $124(4)(1)$ & $0.629(7)(4)\hphantom{0}$ & 
  $0.132(13)(24)$ & $0.101(18)(12)$\\
analysis 2 & $194(7)(1)$ & $123(5)(1)$ & $0.642(10)(4)$ & 
  $0.117(16)(24)$ & $0.093(20)(11)$\\
\bottomrule
\end{tabular}%
\caption{\label{table:results0}Results in the $\MSbar$ scheme at 
$\mu = \unit{2}{\giga\electronvolt}$ from 
the two analysis methods explained in the main text.
The numbers in parentheses denote the statistical error
and our estimate of the uncertainty introduced by 
the renormalization procedure.}
\end{table}%

In table~\ref{table:results0} we compare the results of the two kinds of
final fits that we have performed. The values in the row labelled 
``analysis 1'' have been obtained by fits to all data points, while the 
row labelled ``analysis 2'' contains the results from fits where the 
data with the smallest pion mass have been excluded. 
In the case of $f_\rho^{\vphantom{T}}$, $f_\rho^T$, and 
$f_\rho^T / f_\rho^{\vphantom{T}}$ we have used the fit 
functions \eqref{eq_chpt_fits}, whereas the second 
Gegenbauer moments have been fitted with linear functions of $m_\pi^2$.
One sees that the results of the two fits  are in very good agreement, 
which may be an indication that $\rho$-meson decay, $\rho\to\pi\pi$, 
is not of major importance for the short-distance quantities that we 
are considering here. Discretization errors and finite-size effects 
might be more important, but, unfortunately, cannot be estimated 
reliably using the set of lattices at our disposal. We expect to be 
able to quantify the discretization errors using the new $N_f=2+1$ 
lattice configurations that are generated currently in the framework 
of the CLS initiative~\cite{Bruno:2014jqa}. 

Comparing to the pion case we observe that for the $\rho$ meson we are
able to access the second Gegenbauer moments using momenta with a single
non-zero component (see eqs.~\eqref{eq:amplong} and \eqref{eq:amptrans}),
while we have to consider momenta with two non-vanishing components in
order to compute $a_2$ in the pion. This helps to reduce the statistical
noise and the corresponding error.

As our final results we adopt the numbers from analysis 1. Although
the systematic uncertainty due to finite-volume effects, which we 
cannot estimate reliably, could be sizable in the case of the decay 
constants $f_\rho^T$ and $f_\rho^{\vphantom{T}}$, the agreement with
the experimental value $f_\rho = \unit{210 (4)}{\mega\electronvolt}$ seems 
reasonable. Note that the latter number refers to $\rho^+$ (see 
eq.~\eqref{eq:experiment}). 
Sum rule calculations~\cite{Ball:1996tb,Ball:2006nr} 
yield $f_\rho^{\vphantom{T}} = \unit{206 (7)}{\mega\electronvolt}$ and 
$f_\rho^T = \unit{155 (8)}{\mega\electronvolt}$, where the numbers given 
in ref.~\cite{Ball:2006nr} at the renormalization 
scale $\mu = \unit{1}{\giga\electronvolt}$ have been evolved to 
$\mu = \unit{2}{\giga\electronvolt}$ in leading order of perturbation theory
with $N_f=2$. In view of the fact that the systematics are not yet 
fully controlled, the discrepancies do not look worrying.

In table~\ref{table:results} we compare our main results, i.e., the values
of $f_\rho^T/f_\rho^{\vphantom{T}}$, $a_2^\goodparallel$, and $a_2^\goodperp$,
with QCD sum rule estimates and older lattice data. The statistical and 
renormalization errors of our results have been added in quadrature.  
Again, the sum rule numbers at 
$\mu = \unit{2}{\giga\electronvolt}$ have been obtained from 
the original results at $\mu = \unit{1}{\giga\electronvolt}$ 
by leading order evolution with $N_f=2$. 

Some of these quantities have already been investigated on the lattice.
The BGR collaboration~\cite{Braun:2003jg} has evaluated the ratio
$f_\rho^T/f_\rho^{\vphantom{T}}$ in the quenched approximation with 
chirally improved fermions at a lattice spacing $a=\unit{0.10}{\femto\meter}$
and found $f_\rho^T/f_\rho^{\vphantom{T}} = 0.742(14)$ at 
$\mu = \unit{2}{\giga\electronvolt}$. Further related results have been
reported in refs.~\cite{Capitani:1999zd,Becirevic:2003pn,Gockeler:2005mh}.
The RBC and UKQCD collaborations~\cite{Allton:2008pn} have used $N_f=2+1$
domain-wall fermions at a lattice spacing $a=\unit{0.114}{\femto\meter}$
and masses down to $m_\pi =  \unit{330}{\mega\electronvolt}$
to obtain $f_\rho^T/f_\rho^{\vphantom{T}} = 0.687(27)$ at 
$\mu = \unit{2}{\giga\electronvolt}$. In ref.~\cite{Arthur:2010xf} they 
found $\langle \xi^2 \rangle^{\goodparallel} = 0.27(1)(2)$
at the same scale. Adding the two errors in quadrature and utilizing 
the relation~\eqref{eq:a2xi2} yields $a_2^{\goodparallel}=0.20(6)$. 

\begin{table}[t]
\centering%
\begin{tabular}{llll}%
\toprule
   & \multicolumn{1}{c}{$f_\rho^T/f_\rho^{\vphantom{T}}$} 
& \multicolumn{1}{c}{$a_2^\goodparallel$} 
& \multicolumn{1}{c}{$a_2^\goodperp$} \\
\midrule
this work & $0.629(8)$ & $0.132(27)$ & $0.101(22)$\\
sum rules~\cite{Ball:1996tb,Ball:2006nr} & $0.74(5)$ & $0.11(5)$ & $0.11(5)$\\
lattice~\cite{Becirevic:2003pn} &$0.72(3)$& \mcemd & \mcemd \\
lattice~\cite{Braun:2003jg} &$0.742(14)$& \mcemd & \mcemd \\
lattice~\cite{Allton:2008pn} &$0.687(27)$& \mcemd & \mcemd \\
lattice~\cite{Arthur:2010xf} & \mcemd &$0.20(6)$& \mcemd \\
\bottomrule
\end{tabular}%
\caption{\label{table:results}
Final results together with QCD sum rule estimates and older 
lattice QCD data. The renormalization scale is 
$\mu = \unit{2}{\giga\electronvolt}$.}
\end{table}%

All existing results are, generally, in good agreement, apart from the 
ratio of decay constants $f_\rho^T/f_\rho^{\vphantom{T}}$, which in our case 
is somewhat smaller than the values obtained in other investigations.
This ratio depends strongly on the pion mass, 
cf.~figure~\ref{figure_chiral_extrapolation_norm}, and the extrapolation 
could be affected by the  $\rho\to\pi\pi$ decay at this level of accuracy. 
Clarification of this issue by doing a L{\"u}scher-type analysis 
including four-quark interpolators would be highly desirable since the 
tensor coupling enters the QCD calculations of the $B$-decay form factors
at large recoil (see, e.g., ref.~\cite{Straub:2015ica}), where, in some cases, 
there is a tension with predictions of the Standard Model. Our value for 
the second Gegenbauer coefficient $a_2$ is significantly more precise
compared to previous results. At this level of accuracy, we start 
to be sensitive to the difference between the longitudinally and 
transversely polarized mesons. Our results suggest that $a_2^\goodparallel$ 
may be slightly larger than $a_2^\goodperp$,
although the difference is not yet statistically significant.    
The 20\% accuracy for  $a_2^\goodparallel$ achieved in our work is 
interesting for studies of deeply-virtual vector meson production 
in electron nucleon scattering using the GPD formalism~\cite{Mueller:2013caa}.
Such processes will be investigated with high priority at the JLAB 
$\unit{12}{\giga\electronvolt}$ upgrade and, in the future, at the EIC.

The work reported here will be continued using CLS $N_f=2+1$ lattice 
configurations~\cite{Bruno:2014jqa}. 
Apart from the study of discretization errors our goal is to consider 
DAs of the whole $SU(3)_f$ meson octet, with emphasis on properties of 
the $K^*$ meson, which is of prime importance for flavor physics.
This work is in progress.

\acknowledgments

This work has been supported by the Deutsche Forschungsgemeinschaft 
(SFB/TRR~55), the Studienstiftung des deutschen Volkes, and the 
European Union under the Grant Agreement IRG 256594. The ensembles were
generated primarily on the QPACE systems of the SFB/TRR~55. The 
BQCD~\cite{Nakamura:2010qh} and CHROMA~\cite{Edwards:2004sx} software
packages were used, along with the locally deflated domain decomposition
solver implementation of openQCD~\cite{Luscher:2012av,LuscherOpenQCD}.
We thank Benjamin Gl\"a{\ss}le for software support.
Part of the analysis was also performed on Regensburg's Athene HPC cluster,
the Regensburg HPC-cluster iDataCool and computers of various institutions 
which we acknowledge below. 

One of the authors (JAG) thanks Dr.\ P.E.L.\ Rakow for useful discussions.
Helpful conversations with G.S.\ Bali are gratefully acknowledged.
The work of JAG was carried out with the support of STFC through the
Consolidated Grant ST/L000431/1.

The authors gratefully acknowledge the Gauss Centre for Supercomputing (GCS) 
for providing computing time for a GCS Large-Scale Project on the GCS 
share of the supercomputer JUQUEEN~\cite{juqueen} at J\"ulich Supercomputing 
Centre (JSC) as well as the GCS supercomputer SuperMUC at Leibniz 
Supercomputing Centre (LRZ, www.lrz.de). GCS is the alliance of the three 
national supercomputing centres HLRS (Universit\"at Stuttgart), JSC 
(Forschungszentrum J\"ulich), and LRZ (Bayerische Akademie der 
Wissenschaften), funded by the German Federal Ministry of Education and 
Research (BMBF) and the German State Ministries for Research of 
Baden-W{\"u}rttemberg (MWK), Bayern (StMWFK) and Nordrhein-Westfalen 
(MIWF). 

\clearpage\appendixpage
\begin{appendices}
\section{Transversity operators in the continuum}
\label{app_trans}

In this section we review our construction of the continuum
Green's functions which will be used for connecting the 
$\MSbar$ scheme to the RI${}^\prime$-SMOM scheme employed on the lattice. The
procedure we follow has already been applied to several similar quark bilinear
operators \cite{Gracey:2011fb,Gracey:2011zn,Gracey:2011zg} and we 
will highlight the salient differences for the 
transversity operators considered here. The notation of this section very much 
runs parallel to, for instance, ref.~\cite{Gracey:2011zn}, to which we refer 
the interested reader for more background. First, the two classes of 
operators we are interested in are the flavor non-singlet operators, 
\begin{equation}
\begin{split}
{\cal O}^{T_2}_{\mu\nu\sigma} &= {\cal S} \bar{\psi} \varsigma_{\mu\nu} 
D_\sigma \psi ~~~,~~~ 
{\cal O}^{\partial T_2}_{\mu\nu\sigma} ~=~ {\cal S} \partial_\sigma \left(
\bar{\psi} \varsigma_{\mu\nu} \psi \right) \\
{\cal O}^{T_3}_{\mu\nu\sigma\rho} &= {\cal S} \bar{\psi} \varsigma_{\mu\nu} 
D_\sigma D_\rho \psi ~~,~~
{\cal O}^{\partial T_3}_{\mu\nu\sigma\rho} ~=~ {\cal S} \partial_\sigma \left(
\bar{\psi} \varsigma_{\mu\nu} D_\rho \psi \right) ~~,~~ 
{\cal O}^{\partial\partial T_3}_{\mu\nu\sigma\rho} ~=~ {\cal S} \partial_\sigma
\partial_\rho \left( \bar{\psi} \varsigma_{\mu\nu} \psi \right) ~~~,
\end{split}
\label{opdef}
\end{equation}
where the operators with a single derivative have been 
included for completeness. We define 
$\varsigma^{\mu\nu}$~$=$~$\tfrac{1}{2} [\gamma^\mu,\gamma^\nu]$ which is
related to $\sigma^{\mu\nu}$ by 
\begin{equation}
\sigma^{\mu\nu} ~=~ i \varsigma^{\mu\nu} ~. 
\end{equation} 
Our use of $\varsigma^{\mu\nu}$ is to retain the same conventions with earlier 
renormalization of similar operators 
\cite{Gracey:2011fb,Gracey:2011zn,Gracey:2011zg} and our use of generalized
$\gamma$-matrices which we discuss later. To define the action of the symbol 
${\cal S}$, which imposes certain symmetrization and tracelessness conditions, 
it is best to consider the generalized transversity operators 
${\cal O}^T_{\mu\nu_1\ldots\nu_i\ldots\nu_n}$ from which we will focus on the 
values of $n$~$=$~$2$ and $3$. Specifically, \cite{Hayashigaki:1997dn}, 
\begin{equation}
\eta^{\mu\nu_i}{\cal O}^T_{\mu\nu_1\ldots\nu_i\ldots\nu_n} ~=~ 0 ~~~~
(i ~\geq~ 1) ~~~,~~~ 
\eta^{\nu_i\nu_j}{\cal O}^T_{\mu\nu_1\ldots\nu_i\ldots\nu_j\ldots\nu_n} ~=~
0
\end{equation}
where the label $T$ includes all possible total derivative operators. When
$n$~$=$~$2$, for example, then
\begin{eqnarray}
{\cal S} \bar{\psi} \varsigma^{\mu\nu} D^\sigma \psi &=& \bar{\psi}
\varsigma^{\mu\nu} D^\sigma \psi ~+~ \bar{\psi} \varsigma^{\mu\sigma} D^\nu 
\psi ~-~ 
\frac{2}{(d-1)} \eta^{\nu\sigma} \bar{\psi} \varsigma^{\mu\lambda} D_\lambda 
\psi \nonumber \\
&& +~ \frac{1}{(d-1)} \left( \eta^{\mu\nu} \bar{\psi} \varsigma^{\sigma\lambda}
D_\lambda \psi + \eta^{\mu\sigma} \bar{\psi} \varsigma^{\nu\lambda} 
D_\lambda \psi \right) 
\end{eqnarray}
for the first operator of the $T_2$ sector with again a parallel definition for
the total derivative operator \cite{Gracey:2009da}. In our construction 
for the $T_3$ operators we have taken the convention to include an extra 
factor of $1/6$ in the definition of ${\cal S}$. 
We will use $T_2$ and $T_3$ to refer 
to a sector as well as for the non-total derivative operator of each set. It 
will be clear from the context which is meant. The labelling for each
derivative of a total derivative operator is one $\partial$ symbol applied to 
the sector label. In defining the operators we have omitted the explicit 
flavor indices and note that our perturbative renormalization will be for 
massless quarks; in other words we are in the chiral limit. The total 
derivative operators are required since there is operator mixing within each 
separate sector. It would not usually be necessary to include these but 
since the Green's functions they are needed for are non-forward 
matrix elements, then 
a momentum will flow through the operator insertion and the mixing will be
activated. Part of the evaluation of these matrix elements requires the 
renormalization of the operators. Again our basis choice is partly driven by 
the need to simplify this aspect. Operators with the same quantum numbers and 
dimension will mix under renormalization. However, for our choice the mixing 
matrix will be upper triangular. For instance, we have 
\begin{equation}
{\cal O}^{T_l}_{i\,\mbox{\footnotesize o}} ~=~ \mathcal Z^{T_l}_{ij} {\cal O}^{T_l}_j
\end{equation}
for $l$~$=$~$2$ and $3$ where the subscript ${}_{\mbox{\footnotesize o}}$ 
denotes the bare operator. Then 
\begin{equation}
\mathcal Z^{T_2}_{ij} ~=~ \left(
\begin{array}{cc}
\mathcal Z^{T_2}_{11} & \mathcal Z^{T_2}_{12} \\
0 & \mathcal Z^{T_2}_{22} \\
\end{array}
\right) ~~~,~~~ 
\mathcal Z^{T_3}_{ij} ~=~ \left(
\begin{array}{ccc}
\mathcal Z^{T_3}_{11} & \mathcal Z^{T_3}_{12} & \mathcal Z^{T_3}_{13} \\
0 & \mathcal Z^{T_3}_{22} & \mathcal Z^{T_3}_{23} \\
0 & 0 & \mathcal Z^{T_3}_{33} \\
\end{array}
\right) ~. 
\label{mixmat}
\end{equation}
We use $1$ and $2$ to label the elements of the $T_2$ matrix where $1$ is the
operator $T_2$. Similarly $1$, $2$ and $3$ label the $T_3$ matrix elements 
which respectively correspond to $T_3$, $\partial T_3$ and $\partial \partial
T_3$. The explicit mixing matrix for the $T_2$ system has been determined in 
ref.~\cite{Gracey:2009da} to three loops in the $\MSbar$ scheme. 
Prior to the results we present here, the $T_3$ matrix was known only 
partially to the same order. Entry $(ij)=(11)$ is the 
renormalization constant for the operator $T_3$ itself and the remaining two 
diagonal entries are the same as the operator $T_2$ and the tensor current 
\cite{Broadhurst:1994se,Gracey:2000am,Gracey:2009da}. In other words
the operators of the $T_2$ system without the
total derivatives. In addition the off-diagonal element $(ij)=(23)$ 
is known purely because the non-zero entries of the final two rows 
of $\mathcal Z^{T_3}_{ij}$ are the non-zero entries of the 
$\mathcal Z^{T_2}_{ij}$ matrix. We have determined the final two
off-diagonal elements of $\mathcal Z^{T_3}_{ij}$ by renormalizing 
the operators in a 
quark $2$-point function where the momentum of one of the external quark legs 
is nullified. In other words there is a non-zero momentum flowing through the 
inserted operator. This was the method used to determine a similar mixing 
matrix for the third moment of the usual twist-$2$ Wilson operators in deep 
inelastic scattering \cite{Gracey:2011zg}. However, in 
ref.~\cite{Gracey:2011zg} 
it was noted that such a computational setup was not sufficient to 
determine each of the $(ij)=(12)$ and $(ij)=(13)$ elements separately. 
To disentangle them an extra piece of information 
was required. This is achieved here for $T_3$ by the identity
\begin{equation}
\mathcal Z^{T_3}_{12} ~=~ \mathcal Z^{T_3}_{22} ~-~ \mathcal Z^{T_3}_{11} 
\end{equation}
which is straightforward to establish by integration by parts. Thus to deduce
these remaining two off-diagonal elements we have applied the {\sc Mincer}
algorithm \cite{Gorishnii:1989gt} to the three loop renormalization 
of the operator $T_3$.
As the resulting anomalous dimensions for $T_2$ are given 
in ref.~\cite{Gracey:2009da}, we 
record the first row of the three loop anomalous dimension mixing matrix for 
$T_3$ which is
\begin{eqnarray}
\gamma^{T_3}_{11}(a) &=& \frac{13}{3} C_F a ~+~ 
C_F [1195 C_A - 311 C_F - 452 \Nf T_F ] \frac{a^2}{54} \nonumber \\
&& +~ C_F [10368 \zeta_3 C_A^2 + 126557 C_A^2 - 31104 \zeta_3 C_A C_F 
- 30197 C_A C_F - 67392 \zeta_3 C_A \Nf T_F 
\nonumber \\
&& ~~~~~~~~
- 38900 C_A \Nf T_F + 20736 \zeta_3 C_F^2 - 17434 C_F^2
+ 67392 \zeta_3 C_F \Nf T_F 
\nonumber \\
&& ~~~~~~~~
- 50552 C_F \Nf T_F - 4816 \Nf^2 T_F^2 ] 
\frac{a^3}{972} ~+~ O(a^4) \,, \nonumber \\
\gamma^{T_3}_{12}(a) &=& -~ \frac{4}{3} C_F a ~+~ 
C_F [ - 125 C_A + 34 C_F + 64 \Nf T_F ] \frac{a^2}{27} \nonumber \\      
&& +~ C_F [ - 5184 \zeta_3 C_A^2 - 6790 C_A^2 + 15552 \zeta_3 C_A C_F 
- 18557 C_A C_F + 10368 \zeta_3 C_A \Nf T_F 
\nonumber \\
&& ~~~~~~~~
+ 694 C_A \Nf T_F - 10368 \zeta_3 C_F^2 + 16736 C_F^2
- 10368 \zeta_3 C_F \Nf T_F + 10696 C_F \Nf T_F 
\nonumber \\
&& ~~~~~~~~
+ 752 \Nf^2 T_F^2 ] 
\frac{a^3}{486} ~+~ O(a^4) \,, \nonumber \\
\gamma^{T_3}_{13}(a) &=& -~ \frac{1}{3} C_F a ~+~ 
C_F [11 C_A - 109 C_F + 20 \Nf T_F ] \frac{a^2}{54} \nonumber \\
&& +~ C_F [ - 47952 \zeta_3 C_A^2 + 32969 C_A^2 + 132192 \zeta_3 C_A C_F 
- 138749 C_A C_F
\nonumber \\
&& ~~~~~~~~
+ 25920 \zeta_3 C_A \Nf T_F - 3200 C_A \Nf T_F - 72576 \zeta_3 C_F^2 
+ 27332 C_F^2     
\nonumber \\
&& ~~~~~~~~
- 25920 \zeta_3 C_F \Nf T_F + 39040 C_F \Nf T_F + 2000 \Nf^2 T_F^2 ] 
\frac{a^3}{4860} ~+~ O(a^4) \,,
\label{mixmatt3}
\end{eqnarray}
as the remaining rows are given in ref.~\cite{Gracey:2009da} where
 $a$~$=$~$g^2/(16\pi^2)$. Here
$\zeta_n$ is the Riemann zeta function. We note that our anomalous dimensions 
pass all the usual consistency checks. In particular we derived 
(\ref{mixmatt3}) in an arbitrary linear covariant gauge and checked that the 
gauge parameter cancels as it ought to for gauge invariant operators in the 
$\MSbar$ scheme.  

Having summarized the renormalization of the operators of interest the next
stage is to provide the perturbative corrections to the Green's function where
the operator is inserted in a quark $2$-point function. As we are considering
non-forward matrix elements there is a momentum flowing through the operator. 
More specifically we consider the Green's function $\left\langle 
\psi(p) {\cal O}^{T_l}_{\mu_1 \ldots \mu_{l+1}} (-p-q) \bar{\psi}(q) 
\right\rangle$ for the two cases $l$~$=$~$2$ and $3$. There are two independent
external momenta $p$ and $q$ and we will evaluate the Green's function at the 
fully symmetric point given by  
\begin{equation}
p^2 ~=~ q^2 ~=~ ( p + q )^2 ~=~ -~ \mu^2 \,,
\label{symmpt1}
\end{equation}
from which we have 
\begin{equation}
p \cdot q ~=~ \frac{1}{2} \mu^2 \,,
\label{symmpt2}
\end{equation}
where $\mu$ is a mass scale. For this section we will take this scale to be the
same mass scale that is used in dimensional regularization in 
$d$~$=$~$4$~$-$~$2\epsilon$ dimensions to ensure the coupling constant is 
dimensionless in $d$-dimensions. Therefore, our results for the Green's 
function will not have any logarithms of mass parameter ratios. As each Green's
function has free Lorentz indices we choose to decompose them into a basis of 
Lorentz tensors denoted by ${\cal P}^{T_2}_{(k) \, \mu\nu\sigma}(p,q)$ and
${\cal P}^{T_3}_{(k) \, \mu\nu\sigma\rho}(p,q)$. Here $T_2$ and $T_3$ indicate 
the sector as the basis will be the same for the Green's function with the 
total derivative operators of each sector too. The choice of tensors in each 
basis is not unique. However, each basis is large due to the number of objects
available to build the tensors. These include the momenta $p^\mu$ and 
$q^\mu$ as well as $\eta^{\mu\nu}$. In addition there are Lorentz tensors 
built from the $\gamma$-matrices. As in previous perturbative evaluations 
\cite{Gracey:2011zn,Gracey:2011zg} we use
the generalized $\gamma$-matrices of 
\cite{Kennedy:1981kp,Bondi:1989nq,Vasiliev:1995qj} denoted by
$\Gamma_{(n)}^{\mu_1 \ldots \mu_n}$ and defined by
\begin{equation}
\Gamma_{(n)}^{\mu_1 \ldots \mu_n} ~=~ \gamma^{[\mu_1} \ldots \gamma^{\mu_n]} \,,
\end{equation}
for integers $n$~$\geq$~$0$. In the definition an overall factor of $1/n!$ is
understood. These matrices span the spinor space when dimensional 
regularization is used. As an aside we note that it is in this context that our
choice of $\varsigma^{\mu\nu}$ in the operator definition fits naturally. The 
algebra and properties of these matrices is well-established
\cite{Vasiliev:1996rd,Vasiliev:1996nx}. We 
note one specific property which is important here which is  
\begin{equation}
\mbox{tr} \left( \Gamma_{(m)}^{\mu_1 \ldots \mu_m}
\Gamma_{(n)}^{\nu_1 \ldots \nu_n} \right) ~ \propto ~ \delta_{mn}
I^{\mu_1 \ldots \mu_m \nu_1 \ldots \nu_n} ~,
\label{gamtr}
\end{equation}
where there is no sum over repeated $m$ or $n$ and $I^{\mu_1 \ldots \mu_m \nu_1
\ldots \nu_n}$ is the generalized unit matrix. The key point is that this trace
partitions the space spanned by the tensors in the basis into distinct sectors.
As we consider the operators in massless QCD, only $\Gamma_{(0)}$,
$\Gamma_{(2)}^{\mu\nu}$ and $\Gamma_{(4)}^{\mu\nu\sigma\rho}$ will be needed.
For $T_3$ it might be expected that $\Gamma_{(6)}^{\mu_1\ldots\mu_6}$ would be
required but the symmetrization conditions exclude this $\gamma$-matrix 
from the basis. Finally with these objects we have constructed the 
tensor basis for each sector. For $T_2$ that involves $30$ tensors 
consistent with the symmetry properties of the inserted 
operator. A sample set is presented below.
For $T_3$ there are $42$ tensors and for space reasons these as well as the
full $T_2$ set are given in the attached data file. 

The next step is to compute the coefficients in the decomposition of each
Green's function into their respective tensor basis. In other words we need the
values of the amplitudes $\Sigma^{{\cal O}^{T_l}}_{(k)}(p,q)$ where we write 
\begin{equation}
\left. \left\langle \psi(p) {\cal O}^{T_l}_{\mu_1\ldots\mu_{l+1}}(-p-q)
\bar{\psi}(q) \right\rangle \right|_{p^2 = q^2 = - \mu^2} ~=~ {\cal C}_l
\sum_{k=1}^{N_l} {\cal P}^{T_l}_{(k) \, \mu_1\ldots\mu_{l+1}}(p,q) \, 
\left. \Sigma^{{\cal O}^{T_l}}_{(k)}(p,q) \right|_{p^2 = q^2 = - \mu^2} \,,
\end{equation}
with $N_2$~$=$~$30$ and $N_3$~$=$~$42$. The factor ${\cal C}_l$ is a sector
specific normalization to account for the differing dimensionalities of the 
tensor basis and Green's functions for each sector. Thus we have 
${\cal C}_2$~$=$~$-i$ and ${\cal C}_3$~$=$~$\mu^2$.
The algorithm to determine these 
coefficients has been given in refs.~\cite{Gracey:2011zn,Gracey:2011zg} 
for instance. Briefly, to apply the 
multiloop perturbative integration techniques to find these amplitudes we have 
to extract scalar Feynman integrals which is achieved by a projection method. 
The projection matrix, ${\cal M}^{T_l}_{ij}$, required for each sector is 
constructed from the respective tensor basis
\cite{Gracey:2011zn,Gracey:2011zg} as it is the inverse 
of the matrix
\begin{equation}
{\cal N}^{T_l}_{ij} ~=~ \mbox{tr} \left( \left. {\cal P}^{T_l}_{(i) \, \mu_1 \ldots 
\mu_{l+1}}(p,q) {\cal P}^{{T_l} \, \mu_1 \ldots \mu_{l+1}}_{(j)}(p,q) 
\right|_{p^2=q^2=-\mu^2} \right) ~.
\end{equation}
Due to the size of the matrices, their explicit form is given in the
auxiliary data file provided. Nevertheless,
the partitioning due to the generalized 
$\gamma$-matrices provides a computational shortcut. Hence we have 
\begin{equation}
\Sigma^{{\cal O}^{T_l}}_{(i)}(p,q) ~=~ 
{\cal C}_l^{-1} \sum_{j=1}^{N_l} {\cal M}^{T_l}_{ij} \; \mbox{tr} \left(
{\cal P}^{{T_l} \, \mu_1 \ldots \mu_{l+1}}_{(j)}(p,q) \left. 
\left\langle \psi(p) {\cal O}^{T_l}_{\mu_1 \ldots \mu_{l+1}}(-p-q) 
\bar{\psi}(q) \right\rangle \right|_{p^2=q^2=-\mu^2} \right) ~. 
\end{equation}

Next we briefly note the practical details of actually carrying out the two 
loop evaluation of the Green's function which proceeds in an automatic way. The 
Feynman diagrams are generated using the {\sc Qgraf} 
package \cite{Nogueira:1991ex}. These 
have to be converted to {\sc Form} \cite{Vermaseren:2000nd,Tentyukov:2007mu} 
notation after all the 
Lorentz and color indices have been included. There are $3$ graphs at one 
loop. At two loops there are $32$ graphs for ${\cal O}^{T_2}_{\mu\nu\sigma}$ 
and $37$ for ${\cal O}^{T_3}_{\mu\nu\sigma\rho}$ with fewer graphs for total 
derivative operators. After this the Feynman rules for either operator together
with the propagators and vertices are substituted and the various amplitudes 
are projected out to produce a large number of scalar Feynman integrals 
that need to be calculated. To achieve this we have used the Laporta 
algorithm approach \cite{Laporta:2001dd}. After projection the scalar 
products of the momenta in the 
numerators of the integrals are written in terms of the propagators. In 
addition there may be propagator forms which are not present which are referred
to as irreducible. In this format the Laporta 
algorithm \cite{Laporta:2001dd} is then 
applied which uses integration by parts to systematically construct all the 
algebraic relations between reducible and irreducible scalar integrals for a 
specific momentum topology. The upshot is that all the required scalar 
integrals are written in terms of a small basis of master integrals whose 
$\epsilon$ expansion is known from direct computation 
\cite{Davydychev:1992xr,Usyukina:1992wz,Usyukina:1994iw,Birthwright:2004kk}. 
Therefore, we are able to reduce all the scalar amplitudes to known integrals 
and hence evaluate them exactly at one and two loops. Whilst this is in essence
the Laporta method \cite{Laporta:2001dd} one has to construct the 
relations in a 
practical way. We have chosen to use the {\sc Reduze} package
\cite{Studerus:2009ye}.
Moreover, the output files from the database that {\sc Reduze} builds is 
straightforward to interface with the {\sc Form} modules that constitute the 
automatic computation. For the two loop calculation we perform here, it 
transpires that for the {\sc Reduze} setup there is only one momentum topology 
at one loop and two at two loops. The latter are the ladder and non-planar 
topologies. All the Feynman diagrams that we have to compute can be mapped into
these three cases. The final stage is to carry out the overall renormalization.
This is achieved by computing all the graphs as a function of the bare 
parameters, such as the coupling constant and gauge parameter, following the 
procedure introduced in ref.~\cite{Larin:1993tp} for automatic symbolic 
manipulation loop 
calculations. Then the renormalized parameters are introduced via the usual 
renormalization constant definitions with the operator renormalization 
constants being extracted at the end to leave the finite expressions for each 
scalar amplitude.

To allow orientation to the full data available in the 
attached data file we give a selection of the various amplitudes.
We provide these in numerical form for one representative from each 
$\Gamma_{(n)}$-matrix partition for both operators of the $T_2$ sector. For 
instance, we have 
\begin{align}
\left. \Sigma^{T_2}_{(2)}(p,q) \right| =&
-~ 1.000000 ~+~ [ 0.271008 \alpha + 2.395659 ] a \nonumber \\
& +~ [ 1.329626 \alpha^2  + 2.430759 \alpha - 6.178403 \Nf 
+ 55.151461 ] a^2 ~+~ O(a^3) \,, \nonumber \\
\left. \Sigma^{T_2}_{(23)}(p,q) \right| =&
[ 0.472269 \alpha + 1.416806 ] a \nonumber \\
& +~ [ 1.795895 \alpha^2 + 3.195370 \alpha - 2.817413 \Nf 
+ 36.018151 ] a^2 ~+~ O(a^3) \,, \nonumber \\
\left. \Sigma^{T_2}_{(29)}(p,q) \right| =&
[ - 0.222222 \alpha - 0.666667 ] a \nonumber \\
& +~ [ - 0.808446 \alpha^2 - 4.040708 \alpha + 0.886539 \Nf 
- 14.783322 ] a^2 ~+~ O(a^3) \,, \nonumber \\
\left. \Sigma^{\partial T_2}_{(2)}(p,q) \right| =&
-~ 1.000000 ~+~ [ - 0.062325 \alpha + 0.062325 ] a \nonumber \\
& +~ [ 0.054445 \alpha^2 + 0.640942 \alpha - 1.600114 \Nf 
+ 17.009954 ] a^2 ~+~ O(a^3) \,, \nonumber \\
\left. \Sigma^{\partial T_2}_{(23)}(p,q) \right| =&
[ 0.347245 \alpha + 1.041736 ] a \nonumber \\
& +~ [ 1.302171 \alpha^2 + 3.618039 \alpha - 1.851976 \Nf 
+ 25.400736 ] a^2 ~+~ O(a^3) \,, \nonumber \\
\left. \Sigma^{\partial T_2}_{(29)}(p,q) \right| =&
[ - 1.041737 \alpha - 3.125210 ] a \nonumber \\
& +~ [ - 3.906512 \alpha^2 - 10.854117 \alpha + 5.555928 \Nf 
- 76.202209 ] a^2 ~+~ O(a^3)  \,,
\end{align}
where $\alpha$ is the gauge parameter and the restriction 
$\cdots \left. \vphantom{\Sigma^{T_2}(p,q)} \right|$ stands for evaluation
at (\ref{symmpt1}) and (\ref{symmpt2}). Although we are only interested in the 
values in the Landau gauge, defined by $\alpha$~$=$~$0$, we have performed our 
computations for arbitrary $\alpha$. This is mainly as a check on the 
renormalization of the operators since their anomalous dimensions are 
independent of $\alpha$ in the $\MSbar$ scheme.  

Next we summarize some aspects of the tensor basis and projection
matrix for the $T_2$ sector. Indeed one purpose of this summary is to 
provide an aid to the understanding of the full information given in 
the attached data file for both $T_2$ and $T_3$. Due to the size of the 
bases and matrices we used, a useable electronic format is more appropriate
for their representation.
First, we present a selection of tensors in the $T_2$ basis choosing 
several representatives from each $\Gamma_{(n)}$-matrix partition. 
When one of the external momenta is contracted 
with a Lorentz index then that momentum appears as an index. For example, for 
$T_2$ we have
\begin{eqnarray}
{\cal P}^{T_2}_{(2) \mu \nu \sigma}(p,q) &=&
\varsigma_{\mu \nu} q_\sigma
+ \varsigma_{\mu \sigma} q_\nu
+ \left[ 2 \varsigma_{\mu q} q_\nu q_\sigma
- \varsigma_{\nu q} q_\mu q_\sigma
- \varsigma_{\sigma q} q_\mu q_\nu \right] \frac{1}{\mu^2} \,,
\nonumber \\
{\cal P}^{T_2}_{(5) \mu \nu \sigma}(p,q) &=&
\varsigma_{\nu p} \eta_{\mu \sigma}
+ \varsigma_{\sigma p} \eta_{\mu \nu} \nonumber \\
&& +~ \left[ 2 \varsigma_{\mu p} q_\nu q_\sigma 
+~ d \varsigma_{\nu p} q_\mu q_\sigma 
+ \varsigma_{\nu p} q_\mu q_\sigma 
+ d \varsigma_{\nu q} p_\mu q_\sigma 
+ 3 \varsigma_{\nu q} p_\mu q_\sigma 
+ d \varsigma_{\nu q} p_\sigma q_\mu 
\right. \nonumber \\
&& \left. ~~~~
+~ \varsigma_{\nu q} p_\sigma q_\mu 
+ d \varsigma_{\nu q} q_\mu q_\sigma 
+ 2 \varsigma_{\nu q} q_\mu q_\sigma 
+ d \varsigma_{\sigma p} q_\mu q_\nu 
+ \varsigma_{\sigma p} q_\mu q_\nu 
+ d \varsigma_{\sigma q} p_\mu q_\nu 
\right. \nonumber \\
&& \left. ~~~~
+~ 3 \varsigma_{\sigma q} p_\mu q_\nu 
+ d \varsigma_{\sigma q} p_\nu q_\mu 
+ \varsigma_{\sigma q} p_\nu q_\mu 
+ d \varsigma_{\sigma q} q_\mu q_\nu 
+ 2 \varsigma_{\sigma q} q_\mu q_\nu \right] \frac{1}{\mu^2} \,,
\nonumber \\
{\cal P}^{T_2}_{(17) \mu \nu \sigma}(p,q) &=&
\left[ \varsigma_{\mu p} p_\nu p_\sigma 
- \varsigma_{\mu p} q_\nu q_\sigma 
- \varsigma_{\nu q} p_\mu q_\sigma 
- \frac{1}{2} \varsigma_{\nu q} q_\mu q_\sigma 
- \varsigma_{\sigma q} p_\mu q_\nu 
- \frac{1}{2} \varsigma_{\sigma q} q_\mu q_\nu \right] \frac{1}{\mu^2} \,,
\nonumber \\
{\cal P}^{T_2}_{(23) \mu \nu \sigma}(p,q) &=&
\eta_{\nu \sigma} p_\mu \Gamma_{(0)} \nonumber \\
&& +~ \left[ d p_\mu q_\nu q_\sigma 
-~ \frac{2d}{3} p_\nu p_\sigma q_\mu 
+ \frac{2}{3} p_\nu p_\sigma q_\mu 
- \frac{d}{3} p_\nu q_\mu q_\sigma 
+ \frac{4}{3} p_\nu q_\mu q_\sigma 
\right. \nonumber \\
&& \left. ~~~~
-~ \frac{d}{3} p_\sigma q_\mu q_\nu 
+ \frac{4}{3} p_\sigma q_\mu q_\nu 
+ \frac{d}{3} q_\mu q_\nu q_\sigma 
+ \frac{2}{3} q_\mu q_\nu q_\sigma \right] \frac{\Gamma_{(0)}}{\mu^2} \,,
\nonumber \\
{\cal P}^{T_2}_{(27) \mu \nu \sigma}(p,q) &=&
\left[ 
p_\mu p_\nu p_\sigma 
- p_\mu q_\nu q_\sigma 
- p_\nu q_\mu q_\sigma 
- p_\sigma q_\mu q_\nu 
- q_\mu q_\nu q_\sigma \right] \frac{\Gamma_{(0)}}{\mu^2} \,,
\nonumber \\
{\cal P}^{T_2}_{(29) \mu \nu \sigma}(p,q) &=&
\left[
\Gamma_{(4) \, \mu \nu p q} p_\sigma 
+ \Gamma_{(4) \, \mu \sigma p q} p_\nu \right] \frac{1}{\mu^2} ~.
\label{tenbas}
\end{eqnarray}
We have only shown one tensor from the final partition as the other is given by
replacing the uncontracted vector $p$ by $q$.

For each of the bases we have explicitly constructed the projection matrix 
coefficients. For $T_2$ as there are $30$ projectors this would correspond to a 
$30$~$\times$~$30$ matrix where the entries are rational polynomials in $d$. 
However, as we are using the generalized basis of $\gamma$-matrices in 
$d$-dimensions the projector matrix is block diagonal due to the property of
(\ref{gamtr}). In other words 
\begin{eqnarray}
{\cal M}^{T_2} &=& \left(
\begin{array}{cccc}
{\cal M}^{T_2}_{(2)} & 0 & 0 \\
0 & {\cal M}^{T_2}_{(0)} & 0 \\
0 & 0 & {\cal M}^{T_2}_{(4)} \\
\end{array}
\right) \,,
\end{eqnarray}
where the subscript on the block matrices corresponds to the label of the
analogous $\Gamma_{(n)}$-matrix appearing in the projection tensor. Each of 
these partitions is of different size being respectively $22$, $6$ and $2$
dimensional. Given the size of the first two submatrices it is again not 
feasible to display all entries. Instead we choose to give a few reference 
entries to facilitate the extraction of the full matrices from  
the data file. We have 
\begin{eqnarray}
\mu^2 \, {\cal M}^{T_2}_{(2)~6\,20} &=& -~ \frac{2}{9(d - 2)(d - 3)d} ~~~,~~~
\mu^2 \, {\cal M}^{T_2}_{(2)~15\,10} ~=~ \frac{8(d + 1)}{27(d - 2)d} 
~~~, \nonumber \\
\mu^2 \, {\cal M}^{T_2}_{(0)~3\,6} &=& -~ \frac{2}{9(d - 2)} ~~~,~~~
\mu^2 \, {\cal M}^{T_2}_{(0)~4\,2} ~=~ -~ \frac{1}{3(d - 1)(d - 2)} ~~~,
\end{eqnarray} 
where indices of ${\cal M}^{T_2}_{(0)~i\,j}$ range from $1$ to $6$ and these 
can be mapped to the labels of the tensor basis by adding $22$. Finally, the 
remaining sector is compact enough to record it completely as
\begin{equation}
{\cal M}^{T_2}_{(4)} ~=~ -~ \frac{1}{9(d-2)(d-3)} \left(
\begin{array}{cccc}
2 & 1 \\
1 & 2 \\
\end{array}
\right) ~.
\end{equation}
Overall the matrix ${\cal M}^{T_2}$ is symmetric as is ${\cal M}^{T_3}$. 
Finally, this information should be sufficient to connect with the full
electronic representation for both sectors.

\section{Chiral extrapolation}
\label{app_chpt}

\subsection{Effective field theory framework}

In the specific framework of Chiral Perturbation Theory (ChPT, see, e.g., 
refs.~\cite{Gasser:1983yg,Leutwyler:1993iq,Scherer:2002tk}) applied here, 
the generating functional of all QCD correlators is evaluated by means 
of a path integral involving an {\em effective}\, low-energy Lagrangian 
$\mathcal{L}_{\mathrm{eff}}(U,v,a,s,p,\ldots)$ 
(compare with ref.~\cite{Gasser:1983yg}, and eqs.~(1) and (2) of 
ref.~\cite{Ecker:1989yg}),
\begin{eqnarray}
e^{iZ\lbrack v,a,s,p,\bar{t}\rbrack} &=& \langle 0|\, \mathrm T \exp\biggl(i\int d^{4}x\,\bar{q}\lbrack\gamma_{\mu}(v^{\mu}+\gamma_{5}a^{\mu})-(s-ip\gamma_{5}) + \sigma_{\mu\nu}\bar{t}^{\mu\nu}\rbrack q\biggr)|0\rangle \nonumber \\
 &=&  \int\lbrack dU\rbrack \exp\biggl(i\int d^{4}x\,\mathcal{L}_{\mathrm{eff}}(U,v,a,s,p,\bar{t})\biggr)\,.\label{eq:Zfunctional}
\end{eqnarray}
Formally, all QCD Green's functions can be obtained by taking functional 
derivatives of the generating functional w.r.t.\ the external 
(Hermitian) scalar, pseudoscalar, vector, axial-vector and antisymmetric 
tensor source fields $s,p,v^{\mu},a^{\mu},\bar{t}^{\mu\nu}$. It should be 
noted that the tensor structure with an additional $\gamma_{5}$ is not 
independent due to the identity 
$\sigma_{\mu\nu}\gamma_{5}=\frac{i}{2}\epsilon_{\mu\nu\rho\sigma}\sigma^{\rho\sigma}$. 
The dots stand for other possible source fields (for example, the coupling 
to symmetric tensor fields has been considered in ref.~\cite{Donoghue:1991qv,Dorati:2007bk}). 
The tensor source $\bar{t}^{\mu\nu}$ has been incorporated in ref.~\cite{Cata:2007ns}.
The matrix field $U$ collects the pion (Goldstone boson) fields in a 
convenient way (see below). The effective Lagrangian has to be invariant 
under {\em local}\, chiral transformations of the Goldstone boson and 
source fields, and shares all other symmetries of $\mathcal{L}_{\mathrm{QCD}}$.
A formal proof that low-energy QCD can indeed be analyzed in this way 
has been given by Leutwyler~\cite{Leutwyler:1993iq}.
Under chiral transformations $(L,R)\in SU(2)_{L}\times SU(2)_{R}$, 
the quark and external source fields transform as 
\begin{equation*} \begin{array}{rcl@{$\quad$}rcl}
q_{L}:=\frac{1}{2}(1-\gamma_{5})q & \to & Lq_{L}\,,& 
q_{R}:=\frac{1}{2}(1+\gamma_{5})q & \to & Rq_{R}\,,\\[0.2cm]
l^{\mu} := v^{\mu}-a^{\mu} & \to & 
   Ll^{\mu}L^{\dagger}+iL\partial^{\mu}L^{\dagger}\,,
& r^{\mu} := v^{\mu}+a^{\mu} & \to & 
             Rr^{\mu}R^{\dagger}+iR\partial^{\mu}R^{\dagger}\,,\\[0.2cm]
s+ip & \to&  R(s+ip)L^{\dagger}\,, &
s-ip & \to & L(s-ip)R^{\dagger}\,,\\[0.2cm]
t^{\mu\nu} & \to & Rt^{\mu\nu}L^{\dagger}\,, & {} & {} & {} 
\end{array}
\end{equation*}
where
\begin{eqnarray*}
t^{\mu\nu} &:=& P_{L}^{\mu\nu\rho\sigma}\bar{t}_{\rho\sigma}\,,\quad \bar{t}^{\mu\nu} = P_{L}^{\mu\nu\rho\sigma}t_{\rho\sigma}+P_{R}^{\mu\nu\rho\sigma}t^{\dagger}_{\rho\sigma}\,,\\
P_{L}^{\mu\nu\rho\sigma} &=& \frac{1}{4}\left(g^{\mu\rho}g^{\nu\sigma}-g^{\nu\rho}g^{\mu\sigma}-i\epsilon^{\mu\nu\rho\sigma}\right)\,,\quad P_{R}^{\mu\nu\rho\sigma} = \frac{1}{4}\left(g^{\mu\rho}g^{\nu\sigma}-g^{\nu\rho}g^{\mu\sigma}+i\epsilon^{\mu\nu\rho\sigma}\right)\,.
\end{eqnarray*}

The effective Lagrangian and the perturbative series are ordered by a 
low-energy power counting scheme, counting suppression powers of 
Goldstone boson momenta and masses (or quark masses). For details and 
further references, we refer to refs.~\cite{Gasser:1983yg,Leutwyler:1993iq,Scherer:2002tk}. 
At leading chiral order, the effective Lagrangian describing the 
interaction of the pseudo-Goldstone bosons (pions) with the external source 
fields and each other is given by (see ref.~\cite{Gasser:1983yg})
\begin{equation}\label{eq:L2M}
\mathcal{L}_{M}^{(2)} = \frac{F^{2}}{4}\langle\nabla_{\mu}U^{\dagger}\nabla^{\mu}U\rangle+\frac{F^{2}}{4}\langle \chi U^{\dagger}+U\chi^{\dagger}\rangle\,,
\end{equation}
with $\chi=2B(s+ip)$, $s=\mathcal{M}+\delta s$, where $\mathcal{M}$ is the 
quark mass matrix, and $\delta s$ the remaining part of $s$. The brackets 
$\langle\cdots\rangle$ denote the flavor (or isospin) trace, $F$ is the 
pion decay constant in the chiral limit 
($F\approx \unit{86}{\mega\electronvolt}$), 
and $\nabla_{\mu}U = \partial_{\mu}U-i(v_{\mu}+a_{\mu})U+iU(v_{\mu}-a_{\mu})\,$.
Here $U=\exp(i\sqrt{2}\phi/F)$ with $\phi = \phi^{j}\bar{\lambda}^{j}$, 
where $j$ is a channel (particle species) index which labels the specific 
pion, and $\bar{\lambda}$ are the pertaining channel matrices. We write 
out $\phi$ as
\begin{equation}\label{eq:channelmpi}
\phi = \pi^{0}\bar{\lambda}^{\pi^{0}}+\pi^{+}\bar{\lambda}^{\pi^{+}}+\pi^{-}\bar{\lambda}^{\pi^{-}}\,,\quad \bar{\lambda}^{\pi^{+}} = \frac{1}{2}(\tau^{1}+i\tau^{2})\,,\quad \bar{\lambda}^{\pi^{-}} = \frac{1}{2}(\tau^{1}-i\tau^{2})\,,\quad \bar{\lambda}^{\pi^{0}} = \frac{1}{\sqrt{2}}\tau^{3}\,,
\end{equation}
where the $\tau^{a}$ are the Pauli matrices. The matrix field $U$ 
transforms as $U\rightarrow RUL^{\dagger}$ under chiral transformations. We 
also introduce $u$ as the square root of $U$, $u^2=U$, which transforms 
as $u\rightarrow RuK^{\dagger}=KuL^{\dagger}$, thus defining the so-called 
compensator matrix $K=K(L,R,U)$ (which is also unitary). Below we shall 
set the external fields $p,a$ to zero, $s=\mathcal{M}$ (the quark mass 
matrix) and set $v_{\mu}=v_{\mu}^{a}\frac{\tau^{a}}{2}$, 
$t_{\mu\nu}=t_{\mu\nu}^{a}\frac{\tau^{a}}{2}$\,. At fourth order, we have
\begin{equation}\label{eq:L4M}
\mathcal{L}_{M}^{(4)} = i\frac{\ell_{6}}{4}\langle F^{+}_{\mu\nu}\lbrack u^{\mu},\,u^{\nu}\rbrack\rangle - i\frac{\Lambda_{2}}{2}\langle T^{+}_{\mu\nu}\lbrack u^{\mu},\,u^{\nu}\rbrack\rangle + \cdots\,,
\end{equation}
where we only show the terms needed for our present work (see 
refs.~\cite{Gasser:1983yg,Cata:2007ns,Cata:2009dq} for the complete 
Lagrangian at that order, and eq.~(\ref{eq:resbuildingblocks}) for the 
definition of the operators $u^{\mu}$, $F^{\pm}_{\mu\nu}$ and 
$T^{\pm}_{\mu\nu}$).

\subsection{Chiral Lagrangians for resonances}

Explicit vector meson degrees of freedom have been incorporated in the 
effective Lagrangian of ChPT already in ref.~\cite{Ecker:1989yg,Ecker:1988te}. 
In the following, a ``heavy vector meson'' framework was set 
up~\cite{Jenkins:1995vb,Davoudiasl:1995ed,Bijnens:1997ni,Bijnens:1997rv,
Bijnens:1998di} to deal with problems related to the modified power 
counting in the extended effective theory, caused by introducing a new 
heavy mass scale (the vector meson mass in the chiral limit). Today, 
it is better understood how to deal with such problems in a manifestly 
Lorentz-invariant way, by employing the freedom of choice of the 
subtraction scheme for the effective field 
theory~\cite{Tang:1996ca,Becher:1999he,Fuchs:2003qc}. Such methods have 
been applied to the case of heavy meson resonances in 
refs.~\cite{Borasoy:2001ik,
Fuchs:2003sh,Bruns:2004tj,Bruns:2008ub,Djukanovic:2009zn,Bruns:2013tja,
Djukanovic:2013mka,Zhou:2016saz}. We refer to these references for 
details on the vector meson effective field theory outlined below. 

Keeping in mind the transformation behavior of the external source 
fields $v_{\mu}$, $a_{\mu}$ and $t_{\mu\nu}$ given above, we can write 
down the following terms describing the interaction of the vector mesons 
with the external source fields and the pions (compare also the previous 
references, and ref.~\cite{Faessler:1999de}):
\begin{eqnarray}
\mathcal{L}_{V}^{\mathrm{int}} &=& \frac{f_{V}}{2\sqrt{2}}\langle F^{+}_{\mu\nu}V^{\mu\nu}\rangle + f_{V}^{\bot}\langle T^{+}_{\mu\nu}V^{\mu\nu}\rangle +\frac{ig_{V}}{2\sqrt{2}}\langle\lbrack u_{\mu},\,u_{\nu}\rbrack V^{\mu\nu}\rangle \nonumber \\ &+& \frac{f_{\pi\omega}}{4\sqrt{2}}\epsilon_{\mu\nu\rho\sigma}\langle V^{\mu}\lbrace u^{\nu},\,F_{+}^{\rho\sigma}\rbrace\rangle + \frac{f_{\pi\omega}^{\bot}}{\sqrt{2}}\epsilon_{\mu\nu\rho\sigma}\langle V^{\mu}\lbrace u^{\nu},\,T_{+}^{\rho\sigma}\rbrace\rangle + \cdots\,,\label{eq:LV}\\
\mathcal{L}_{VV}^{\mathrm{int}} &=& \frac{g_{A}^{V}}{2}\epsilon_{\mu\nu\rho\sigma}\langle\lbrace D^{\mu}V^{\nu},\,V^{\rho}\rbrace u^{\sigma}\rangle + \cdots\,,\label{eq:LVV}
\end{eqnarray}
where
\begin{eqnarray}
V^{\mu\nu} &=& D^{\mu}V^{\nu}-D^{\nu}V^{\mu}:=\partial^{\mu}V^{\nu}-\partial^{\nu}V^{\mu} + \lbrack\Gamma^{\mu},\,V^{\nu}\rbrack - \lbrack\Gamma^{\nu},\,V^{\mu}\rbrack\,,\nonumber \\
\Gamma^{\mu} &=& \frac{1}{2}\left(u^{\dagger}[\partial^{\mu}-i(v^{\mu}+a^{\mu})]u + u[\partial^{\mu}-i(v^{\mu}-a^{\mu})]u^{\dagger}\right)\,,\nonumber \\
F^{\pm}_{\mu\nu} &=& uF^{L}_{\mu\nu}u^{\dagger} \pm u^{\dagger}F^{R}_{\mu\nu}u\,,\quad u_{\mu}=iu^{\dagger}\left(\nabla_{\mu}U\right)u^{\dagger}\,,\nonumber \\
F^{R,L}_{\mu\nu} &=& \partial_{\mu}(v_{\nu}\pm a_{\nu})-\partial_{\nu}(v_{\mu}\pm a_{\mu})-i\lbrack(v_{\mu}\pm a_{\mu}),\,(v_{\nu}\pm a_{\nu})\rbrack\,, \nonumber \\
T^{\pm}_{\mu\nu} &=& u^{\dagger}t_{\mu\nu}u^{\dagger} \pm ut_{\mu\nu}^{\dagger}u\,,\nonumber \\
\nabla_{\mu}U &=& \partial_{\mu}U-i(v_{\mu}+a_{\mu})U+iU(v_{\mu}-a_{\mu})\,,\quad u=\sqrt{U}\,,\nonumber \\
V_{\mu} &=& \rho^{0}_{\mu}\bar{\lambda}^{\rho^{0}} + \rho^{+}_{\mu}\bar{\lambda}^{\rho^{+}} + \rho^{-}_{\mu}\bar{\lambda}^{\rho^{-}} + \frac{\omega_{\mu}}{\sqrt{2}}\mathds{1}_{2\times 2}\,,\quad \bar{\lambda}^{\rho^{0}}=\bar{\lambda}^{\pi^{0}}\,,\quad \bar{\lambda}^{\rho^{\pm}}=\bar{\lambda}^{\pi^{\pm}}\,, \label{eq:resbuildingblocks} 
\end{eqnarray}
see eq.~(\ref{eq:channelmpi}) for the channel matrices $\bar{\lambda}$. 
We have used a large-$N_{c}$ argument here to cast the $\rho$ and $\omega$ 
fields in the matrix form of the last line in eq.~(\ref{eq:resbuildingblocks}),
compare also with eq.~(27) of ref.~\cite{Jenkins:1995vb}. The dots indicate 
terms of higher chiral order, terms involving external source fields $s,p$ 
(which are not needed here), or terms involving more derivatives, which 
result in contributions of the same form as those resulting from the terms 
given above, when using the equations of motion or field transformations \cite{Fearing:1999fw}.
The vector field propagator in momentum space is
\begin{equation}
D_{\mu\nu}(q) = (-i)\frac{\eta_{\mu\nu}-\frac{q_{\mu}q_{\nu}}{m_V^{2}}}{q^2-m_V^{2}}\,.
\end{equation}

\subsection{Extrapolation formulae}

For the sake of completeness, we first discuss the pion matrix elements
\begin{eqnarray}
\langle 0|\bar{q}\frac{\tau^{a}}{2}\gamma_{\mu}q|\pi^{b}(p)\pi^{c}(k-p)\rangle &=& i\epsilon^{abc}(2p-k)_{\mu}f_{\pi\pi}^{v}(k^2)\,,\\ 
\langle 0|\bar{q}\frac{\tau^{a}}{2}\sigma_{\mu\nu}q|\pi^{b}(p)\pi^{c}(k-p)\rangle &=& \epsilon^{abc}(k_{\mu}p_{\nu}-k_{\nu}p_{\mu})f_{\pi\pi}^{t}(k^2)\,.
\end{eqnarray}
The standard framework of ChPT yields
\begin{align}
f_{\pi\pi}^{v}(k^2) =& 1 -\frac{k^2}{6(4\pi F)^2}\left(96\pi^2\ell_{6}^{r}+\frac{1}{3} +\log\left(\frac{M^2}{\mu_\chi^2}\right)\right) + \frac{4M^2-k^2}{6F^2}\bar{I}_{\pi\pi}(k^2) + O(p^4)\,,\\
f_{\pi\pi}^{t}(k^2) =& \frac{\Lambda_{2}}{F^2}\left(1-\frac{k^2+3M^2}{6(4\pi F)^2}\log\left(\frac{M^2}{\mu_\chi^2}\right) - \frac{k^2}{18(4\pi F)^2} + \frac{4M^2-k^2}{6F^2}\bar{I}_{\pi\pi}(k^2)\right) \nonumber \\ &{} +  \lambda^{r}_{m}M^2+\lambda^{r}_{k}k^2 + O(p^4)\,,
\end{align}
where the loop function $\bar{I}_{\pi\pi}(k^2)$ is given at the end of 
this appendix, in eq.~(\ref{eq:Ipipibar}) (it vanishes for $k^2\rightarrow 0$, 
and is complex for $k^2>4m_\pi^{2}$), and $\ell_{6}^{r},\lambda^{r}_{m},\lambda^{r}_{k}$ 
are renormalized low-energy constants, which depend on the scale 
$\mu_\chi$. $M$ is the leading term in the quark-mass expansion of the 
pion mass $m_\pi$, derived from the Lagrangian (\ref{eq:L2M}) (at the order 
we are working, it can be set equal to the pion mass). We note that, up to 
corrections of two loop order, these expressions for the form factors are 
consistent with the constraints from elastic unitarity, 
\begin{eqnarray}
\operatorname{Im}f_{\pi\pi}^{v}(k^2) &=& f_{\pi\pi}^{v}(k^2)\sigma(k^2)t_{1}^{1\ast}(k^2)\,, \nonumber \\ \operatorname{Im}f_{\pi\pi}^{t}(k^2) &=& f_{\pi\pi}^{t}(k^2)\sigma(k^{2})t_{1}^{1\ast}(k^{2})\,,\quad 4m_\pi^{2}<k^2<16m_\pi^{2}\,,
\end{eqnarray}
where $\sigma(s):=\sqrt{1-\frac{4m_\pi^{2}}{s}}$, and $t_{1}^{1}(s)$ is 
the isospin $I=1$, $p\,$-partial wave amplitude for $\pi\pi$ scattering, 
\begin{displaymath}
t_{1}^{1}(s) = \frac{e^{2i\delta_{1}^{1}(s)}-1}{2i\sigma(s)}\quad\mathrm{for}\quad 4m_\pi^{2}<s<16m_\pi^{2}\,.
\end{displaymath}
It easily follows that the form factors $f_{\pi\pi}^{v,t}$ must have the 
phase $\delta_{1}^{1}(s)$ in the elastic region.

\subsection{\texorpdfstring{Contributions to $\rho$ matrix elements}{Contributions to rho matrix elements}} 
Here we use the definitions
\begin{eqnarray}
\langle 0|\bar{q}\frac{\tau^{a}}{2}\gamma_{\mu}q|\rho^{b}(k,\lambda)\rangle &=& \delta^{ab}m_{\rho}e_{\mu}^{(\lambda)}f_{\rho}/\sqrt{2}\,, \\ 
\langle 0|\bar{q}\frac{\tau^{a}}{2}\sigma_{\mu\nu}q|\rho^{b}(k,\lambda)\rangle &=& i\delta^{ab}(e_{\mu}^{(\lambda)}k_{\nu}-e_{\nu}^{(\lambda)}k_{\mu})f_{\rho}^{T}/\sqrt{2}\,,
\end{eqnarray}
and find at the one loop level up to $O(p^4)$
\begin{align}
\frac{m_{\rho}f_{\rho}}{\sqrt{2}} &= \sqrt{Z^\chi_\rho}\left(f_{V}k^2\left(1-\frac{I_{\pi}}{F^2}\right) + c_{V}k^2M^2 +\frac{4g_{V}f_{\pi\pi}^{v}(k^2)}{F^2}k^2I_{\pi\pi}^{A}(k^2) - \frac{4f_{\pi\omega}g_{A}^{V}}{F^2}k^2I_{\pi\omega}^{A}(k^2)\right)\,,  \label{chiral_f_long}\\
\frac{f_{\rho}^{T}}{\sqrt{2}} &= \sqrt{Z^\chi_\rho}
\Bigg(\frac{f_{V}^{T}}{\sqrt{2}}\left(1-\frac{I_{\pi}}{2F^2}\right)+ \frac{c_{V}^{T}}{\sqrt{2}}M^2 + \frac{2g_{V}f_{\pi\pi}^{t}(k^2)}{F^2}k^2I_{\pi\pi}^{A}(k^2) \nonumber \\
& \qquad \qquad {}
- \frac{4\left(f_{\pi\omega}^{T}+\frac{f_{V}^{T}}{\sqrt{2}}\right)g_{A}^{V}}{F^2}I_{\pi\omega}^{A}(k^2)\Bigg)\,. \label{chiral_f_trans}
\end{align}
Here we have to set $k^2\equiv s$ equal to the rho pole, 
$k^2\rightarrow s_{\mathrm {pole}} = m_{\rho}^{2}-im_{\rho}\Gamma_{\rho}$ 
\cite{Gegelia:2009py}. The wave function renormalization factor is derived 
from the $\rho$ self-energy $\Pi_{\rho}(s)$,
\begin{equation}
Z^\chi_\rho=\frac{1}{1-\frac{d}{ds}\Pi_{\rho}(s)}\biggr|_{s_{\mathrm {pole}}} \approx 1+\frac{d}{ds}\Pi_{\rho}(s)\bigr|_{s_{\mathrm {pole}}}\,,
\end{equation}
where the contribution of the one loop graphs to the self-energy is given 
by (compare ref.~\cite{Bruns:2013tja})
\begin{equation}\label{eq:PirhoLoop}
\Pi_{\rho}^{\mathrm{loop}}(s) := -\frac{4s^2g_{V}^{2}}{F^4}I_{\pi\pi}^{A}(s) - \frac{8s(g_{A}^{V})^{2}}{F^2}I_{\pi\omega}^{A}(s) + \mathrm{tadpoles} \,.
\end{equation}
The ``tadpole'' terms can be taken to be energy independent at the order 
we are working to. The integral $I_{\pi\pi}^{A}$ can be deduced from 
eqs.~(\ref{eq:IpipiDef})--(\ref{eq:IpipiA}) below, and $I_{\pi\omega}^{A}$ 
is given by eq.~(\ref{eq:IMVA}) (with 
$m_V\rightarrow m_{\omega}\approx m_{\rho}$). The local terms 
proportional to $c_{V}$, $c_{V}^{T}$ can be associated with local 
operators $\langle F^{+}_{\mu\nu}\chi_{+}V^{\mu\nu}\rangle$, 
$\langle T^{+}_{\mu\nu}\chi_{+}V^{\mu\nu}\rangle$ etc., and can be 
used to absorb (real) terms of $O(M^2)$ from the loop integrals. 
The loop functions are given at the end of this appendix 
($I_{\pi\omega}^{A}=I_{\pi V}^{A}(m_V\rightarrow m_{\omega}\approx m_{\rho})$).
In the one loop approximation, we evaluate the loop integrals at 
$k^2=m_{\rho}^{2}$ (the imaginary part of the pole position is generated 
by loop graphs). The leading non-analytic term in $I_{\pi\omega}^{A}$ 
is given by
\begin{equation}
I_{\pi\omega}^{A}(m_{\rho}^{2}\approx m_{\omega}^{2}) = \frac{M^3}{48\pi m_{\rho}}+\ldots\,,
\end{equation}
and the terms of order $M^{0}$ and $M^{2}$ are absorbed in the corresponding 
LECs. The chiral logarithm of this integral is of $O(M^4)$. One finds
\begin{equation}\label{eq:refrho}
\mathrm{Re}\,\frac{f_{\rho}^{T}}{f_{\rho}}  \approx \frac{f_{V}^{T}}{\sqrt{2}m_{\rho}f_{V}}\left(1+\frac{M^2}{32\pi^2F^2}\log\left(\frac{M^2}{\mu_\chi^2}\right) + \delta c M^2 - \frac{g_{A}^{V}}{12\pi F^2 m_{\rho}}\left(1+\frac{\sqrt{2}f_{\pi\omega}^{T}}{f_{V}^{T}} - \frac{f_{\pi\omega}}{f_{V}}\right)M^3\right)\,,
\end{equation}
where $\delta c$ is the following combination of (renormalized) LECs,
\begin{equation*}
\delta c:=\frac{c_{V}^{T}}{f_{V}^{T}}-\frac{c_{V}}{f_{V}}\,.
\end{equation*}
The coefficient of the leading chiral logarithm is in agreement with ref.~\cite{Cata:2009dq}.
With $g_{A}^{V}\approx\frac{3}{4}$ (see ref.~\cite{Bruns:2013tja}, and 
references therein), the coefficient of the third-order term should be 
of order $\sim \unit{3}{\power{\giga\electronvolt}{-3}}$. Inserting this estimate, and 
$\mu_\chi=\unit{770}{\mega\electronvolt}$, the third-order term becomes 
comparable to the leading chiral logarithm for $M\gtrsim \unit{200}{\mega\electronvolt}$, 
so it might give a non-negligible contribution for most data points. 

In eq.~(\ref{eq:refrho}), we have written the result for the chiral 
expansion of $\mathrm{Re}\,f_{\rho}^{T}/f_{\rho}$, which motivates the
extrapolation formula \eqref{eq_chpt_fits_ratio}, while the formulae
\eqref{eq_chpt_fits_long} and \eqref{eq_chpt_fits_trans} result from
\eqref{chiral_f_long} and \eqref{chiral_f_trans}, respectively, upon
inserting the explicit expressions for the loop functions given below.
The cusp effects and imaginary parts in the chiral behavior of the 
couplings could only be extracted indirectly from the computed correlators, 
which are real on Euclidean lattices with a finite volume. A 
more thorough analysis is needed to deal with such complications. 
It is, however, important to note that 
the leading non-analytic terms given above are not afflicted by this 
deficiency. This can be deduced from the fact that they agree with the 
corresponding results in the heavy vector meson framework 
\cite{Bijnens:1998di,Cata:2009dq}, where the unitarity effects due to 
the $\pi\pi$ loops are either dropped or derived from contact terms of a 
non-Hermitean part of the effective Lagrangian 
(see, e.g., ref.~\cite{Davoudiasl:1995ed}).

In the expression for the ratio given in eq.~\eqref{eq:refrho}, the factors 
of $\sqrt{Z^\chi_\rho}$ and the non-analytic terms in the loop function 
$I_{\pi\pi}^{A}$ containing the imaginary part cancel at one-loop order. 
Due to this simplification, it is 
straightforward to compute the finite-volume corrections for this ratio. 
Here, we attempt only an estimate of the leading finite-volume correction, 
related to the $\mathcal{O}(M^2)$ `chiral-log' term contained in the 
tadpole loop integral $I_{\pi}$ (compare eq.~\eqref{eq:tadpi} below).
According to the standard formalism of ChPT in a finite cubic volume 
$V=L^3$ \cite{Hasenfratz:1989pk}, this loop integral is replaced by 
its finite-volume counterpart
\begin{equation}\label{eq:tadpifv}
I_{\pi}^{(L)} = I_{\pi} + \sum_{\mathbf{0}\not=\mathbf{k}\in\mathds{Z}^{3}}\frac{MK_{1}\left(ML|\mathbf{k}|\right)}{4\pi^2L|\mathbf{k}|}\,.
\end{equation}
Here $K_{1}(z)$ is the modified Bessel function of the second kind, which 
decays exponentially for large positive $z$, 
$K_{1}(z)\rightarrow\sqrt{\frac{\pi}{2z}}e^{-z}\,$. 
Inserting (\ref{eq:tadpifv}) in \eqref{chiral_f_long} and 
\eqref{chiral_f_trans} yields the leading finite-volume correction to 
the ratio of eq.~(\ref{eq:refrho}) upon a straightforward chiral expansion.

\subsection{Loop functions} 
To render this appendix self-contained, we give the definitions of the 
loop integrals occuring in the formulae above.  
The loop integral with two pion propagators is given by
\begin{equation}\label{eq:IpipiDef}
I_{\pi\pi}(s) = \int\frac{d^{d}l}{(2\pi)^{d}}\frac{i}{((k-l)^2-M^2)(l^2-M^2)}\,,\quad k^2=:s\,.
\end{equation}
It diverges when the space-time dimension $d$ approaches 4,
\begin{align}
I_{\pi\pi}(0) &= 2\bar{\lambda}+\frac{1}{16\pi^2}\left(1+\log\left(\frac{M^2}{\mu_\chi^2}\right)\right)+O(4-d)\,, \nonumber \\
\bar\lambda &= \frac{\mu_\chi^{d-4}}{16\pi^{2}}\biggl(\frac{1}{d-4}-\frac{1}{2}[\log(4\pi)+\Gamma'(1)+1]\biggr) \,, \nonumber
\end{align}
however the difference $\bar{I}_{\pi\pi}(s):=I_{\pi\pi}(s)-I_{\pi\pi}(0)$ 
is finite,
\begin{equation}
\bar{I}_{\pi\pi}(s) = -\frac{s}{16\pi^2}\int_{4M^2}^{\infty}ds'\frac{\sqrt{1-\frac{4M^2}{s'}}}{s'(s'-s)}\,,
\end{equation}
where it is understood that real values of $s$ are approached from the 
upper complex plane for $s\in\lbrack 4M^2,\infty )$. Explicitly,
\begin{equation}\label{eq:Ipipibar}
\bar{I}_{\pi\pi}(s) = -\frac{1}{8\pi^2}\left(1+\sigma_0(s)\,\mathrm{artanh}\left(-\frac{1}{\sigma_0(s)}\right)\right)\,,\quad \sigma_0(s):=\sqrt{1-\frac{4M^2}{s}}\,.
\end{equation}
In the chiral limit ($M\rightarrow 0$),
\begin{displaymath}
I_{\pi\pi}(s)\rightarrow 2\bar{\lambda}-\frac{1}{16\pi^2}\left(1+\log\left(-\frac{\mu_\chi^2}{s}\right)\right)\,.
\end{displaymath}
Note that this integral has an imaginary part 
$\mathrm{Im}\,\bar{I}_{\pi\pi}(s)=-\sigma_0(s)\theta(s-4M^2)/(16\pi)$ 
for real $s>4M^2$. 
We have also employed the abbreviation
\begin{equation}\label{eq:IpipiA}
I_{\pi\pi}^{A}:=\frac{1}{4(d-1)}\left(2I_{\pi}-(s-4M^2)I_{\pi\pi}\right)\,,
\end{equation}
where
\begin{equation}\label{eq:tadpi}
I_{\pi} := \int\frac{d^{d}l}{(2\pi)^{d}}\frac{i}{l^{2}-M^{2}}=2M^{2}\bar{\lambda}+\frac{M^{2}}{16\pi^{2}}\log\left(\frac{M^{2}}{\mu_\chi^{2}}\right)\,,
\end{equation}
for $d\rightarrow 4$. 
The scalar integral including two different propagators can be written as
\begin{align}\label{eq:IpiV}
I_{\pi V}(k^{2}\equiv s) = \int\frac{d^{d}l}{(2\pi)^{d}}\frac{i}{((k-l)^{2}-m_V^{2})(l^{2}-M^{2})} = I_{\pi V}(m_V^{2})-\frac{(s-m_V^{2})}{16\pi^{2}}J^{\pi V}(s)\,,
\end{align}
and we refer to appendix~B of ref.~\cite{Bruns:2013tja} for details on 
the chiral expansion. We also use 
\begin{equation}\label{eq:IMVA}
I_{\pi V}^{A}=\frac{1}{4s(d-1)}\big((4sM^2-(s+M^2-m_V^2)^2)I_{\pi V}+(s+M^2-m_V^2)I_{\pi}
+(s-M^2+m_V^2)I_{V}\big)\,,
\end{equation}
where $I_{V}$ is given by the formula for $I_{\pi}$ with $M\rightarrow m_V$. 
Here, the letter $V$ stands for the vector meson running in the 
loop ($\rho,\omega,\ldots$).

\end{appendices}
\clearpage
\providecommand{\href}[2]{#2}\begingroup\raggedright\endgroup
\end{document}